\documentclass[times,singlecolumn,final]{elsarticle}

\usepackage{medima}
\usepackage{framed,multirow}
\usepackage{booktabs}
\usepackage{amsfonts} 
\usepackage{amssymb}
\usepackage{newtxtext,newtxmath} 
\usepackage{latexsym}
\usepackage{subcaption}
\usepackage{nth}
\usepackage{pgfplots}
\usepackage{pgfplotstable}
\pgfplotsset{compat=1.7}
\usepackage{tikz}

\usepackage{lineno}

\newcommand{\beginsupplement}{%
        \setcounter{table}{0}
        \renewcommand{\thetable}{S\arabic{table}}%
        \setcounter{figure}{0}
        \renewcommand{\thefigure}{S\arabic{figure}}%
     }

\usepackage{url}
\usepackage{xcolor}

\usepackage{hyperref}
\hypersetup{
  colorlinks   = true, 
  urlcolor     = red, 
  linkcolor    = blue, 
  citecolor   = blue 
}

\definecolor{newcolor}{rgb}{.8,.349,.1}
\journal{Medical Image Analysis}
\begin{document}

\verso{Vesal \textit{et~al.}}
\begin{frontmatter}

\title{Domain Generalization for Prostate Segmentation in Transrectal Ultrasound Images: A Multi-center Study \tnoteref{tnote1}}%

\author[1]{Sulaiman \snm{Vesal}\corref{cor1}}
\cortext[cor1]{Corresponding author.}
\author[2]{Iani \snm{Gayo}}
\author[1,3]{Indrani \snm{Bhattacharya}}
\author[4]{Shyam \snm{Natarajan}}
\author[4]{Leonard S. \snm{Marks}}
\author[2]{Dean C \snm{Barratt}}
\author[1]{Richard E. \snm{Fan}}
\author[2]{Yipeng \snm{Hu}}
\author[1]{Geoffrey A. \snm{Sonn}\corref{cor2}}
\author[3]{Mirabela \snm{Rusu}\corref{cor2}}

\ead{svesal@stanford.edu (Sulaiman Vesal), mirabela.rusu@stanford.edu}
\cortext[cor2]{Equally contributed as senior authors.}

\address[1]{Department of Urology, Stanford University, 300 Pasteur Drive, Stanford, CA 94305, USA}
\address[2]{Centre for Medical Image Computing, Wellcome/EPSRC Centre for Interventional \& Surgical Sciences, and Department of Medical Physics \& Biomedical Engineering, University College London, 66-72 Gower St, London WC1E 6EA, UK}
\address[3]{Department of Radiology, Stanford University, 300 Pasteur Drive, Stanford, CA 94305, USA}
\address[4]{Department of Urology, University of California Los Angeles, 200 Medical Plaza Driveway, Los Angeles, CA 90024, USA}

\begin{abstract}
Prostate biopsy and image-guided treatment procedures are often performed under the guidance of ultrasound fused with magnetic resonance images (MRI). Accurate image fusion relies on accurate segmentation of the prostate on ultrasound images. Yet, the reduced signal-to-noise ratio and artifacts (e.g., speckle and shadowing) in ultrasound images limit the performance of automated prostate segmentation techniques and generalizing these methods to new image domains is inherently difficult. In this study, we address these challenges by introducing a novel 2.5D deep neural network for prostate segmentation on ultrasound images. Our approach addresses the limitations of transfer learning and finetuning methods (i.e., drop in performance on the original training data when the model weights are updated) by combining a supervised domain adaptation technique and a knowledge distillation loss. The knowledge distillation loss allows the preservation of previously learned knowledge and reduces the performance drop after model finetuning on new datasets. Furthermore, our approach relies on an attention module that considers model feature positioning information to improve the segmentation accuracy. We trained our model on 764 subjects from one institution and finetuned our model using only ten subjects from subsequent institutions. We analyzed the performance of our method on three large datasets encompassing  2067 subjects from three different institutions.
Our method achieved an average Dice Similarity Coefficient (Dice) of $94.0\pm0.03$ and Hausdorff Distance (HD95) of 2.28 $mm$ in an independent set of subjects from the first institution. Moreover, our model generalized well in the studies from the other two institutions (Dice: $91.0\pm0.03$; HD95: 3.7 $mm$ and Dice: $82.0\pm0.03$; HD95: 7.1 $mm$). We introduced an approach that successfully segmented the prostate on ultrasound images in a multi-center study, suggesting its clinical potential to facilitate the accurate fusion of ultrasound and MRI images to drive biopsy and image-guided treatments.

\end{abstract}


\end{frontmatter}

\section{Introduction}
Prostate cancer is the third most common cancer diagnosed globally and the fifth leading cause of cancer-related mortality and morbidity in men \citep{globalstat}. Transrectal ultrasound-guided (TRUS) biopsy procedures are used to diagnose prostate cancer \citep{MICHALSKI20161038, Sarkar2016}. TRUS images show the prostate in real-time, allowing urologists to guide the biopsy needle during the procedure. Yet, these images have a low signal-to-noise ratio and artifacts (e.g., speckle and shadowing), which reduce the ability of clinicians to reliably distinguish cancerous from normal regions. Therefore, TRUS-guided biopsy procedures usually involve blind sampling of 12 regions throughout the prostate \citep{Harvey2012}. Such blinded systematic biopsy procedures sample <1\% of the prostate, missing 52\% clinically significant cancer \citep{Williams2022, Schimmoller2016}. 
 
To address the inability of TRUS images to reliably show cancer, approaches have been developed to fuse magnetic resonance images (MRI) with TRUS images to project suspicious lesions from MRI onto TRUS images and target them during biopsy \citep{diagnostics11020354, Liau2019}. Fusion requires the registration of MRI and TRUS images which is typically done by aligning the prostate boundary. The prostate boundaries are usually manually outlined on both MRI and TRUS images by clinical experts, usually radiologists for MRI and urologists for TRUS \citep{diagnostics11020354}. Prostate segmentation on TRUS images could also be useful for other applications, including computer-aided diagnosis and targeting cancers using ultrasound images alone.. Accurate manual segmentation of the prostate on TRUS images is a tedious, time-consuming task that suffers from inter- and intra-observer variability due to the reduced quality of the images. Moreover, it is common for the prostate to require segmentation multiple times throughout the biopsy procedure to account for motion and tissue changes \citep{wang2019}. 

Methods for automated prostate segmentation on TRUS images hold the potential to improve accuracy, reduce inter-reader variability and reduce the time required for manual prostate segmentation during clinical procedures. Numerous automated and semi-automated algorithms have been presented for prostate segmentation on TRUS images. Some approaches used shape statistics as prior knowledge \citep{li2016, GHOSE2012262}, yet required the intervention of an expert user. Other approaches extracted textural features from TRUS images and combined them with traditional machine learning methods to formulate a classification task \citep{Zhan2006, Yang2016}. These methods used hand-crafted features for segmentation, which are inadequate for capturing high-level semantic information and failed to deliver accurate segmentation for complex prostate cases. 

Deep learning-based methods have achieved high accuracy in medical image segmentation tasks compared to non-learning methods \citep{Liu2020, Azizi2018, Aldoj2020, vesal2020}, and have already been used for prostate segmentation on TRUS images \citep{Ans2017, Ghavami2018, vanSloun2021, Jaouen2019, Girum2020, wang2019, Nathan2020, Lei2019, Xu2021}. Most deep learning-based segmentation methods rely on supervised encoder-decoder architectures. Some studies incorporated prior shape information as statistical shape models to improve the segmentation of challenging regions, e.g., apex and base \citep{Zeng2018, KARIMI2019186, yang2017}. Other studies explored temporal information of TRUS scans using recurrent neural networks (RNNs) \citep{ANAS2018107}, attention mechanism \citep{wang2019} or shadow augmentation \citep{Xu2021} to improve segmentation quality. However, most studies evaluated their methods on a small set of patients with data from a single institution and a single manufacturer, thus providing limited evidence about generalization across data from other institutions and different imaging devices, vendors, and data acquisition parameters (end-firing and side-firing probes).A recent study \citep{vanSloun2021} investigated the use of deep learning models for multi-center prostate gland segmentation on TRUS images. The author employed a standard UNet architecture to segment the prostate gland. The model was trained and tested on a small number of patients from three different institutions (a total of 436 TRUS images from 181 men), acquired using only end-fire probes. The reduced training population size and homogeneous data possibly limit the generalizability of their approach, particularly in ultrasound images acquired with a different type of probe or at a different institution.

Deep learning models often require large amounts of data during training to achieve robust and consistent performance across data from multiple institutions. Transfer learning or finetuning \citep{Weiss2016} is one method for improving the generalization of segmentation models on new data. The main drawback of classical finetuning approach is that it forgets previous knowledge since the model's weights are being updated \citep{lkd2019, CL_Mib}. As a result, the performance deteriorates when the newly trained model is tested on the previous data \citep{lkd2019}. Knowledge distillation techniques \citep{Lin2022} have been widely used to preserve the high performance of a model when applied to new tasks. It was originally used to retain the performance of a complex model when adopting to a smaller network for more efficient deployment \citep{hinton2015}.

Several studies attempted to apply the knowledge distillation technique for a variety of objectives in the computer vision and medical domains, including cross-modality learning \citep{tian2019crd} (transfer knowledge from one modality to another modality without the need of any additional annotations), metric learning \citep{park2019relational, kim2021selfreg} (map the input feature representations to an embedding space), and network regularization \citep{Yun_2020_CVPR} (enhance the generalization performance of deep neural networks using regularization losses). Similarly, approaches based on knowledge distillation have been developed for domain adaptation \citep{meng2018, zhou2020}. The domain adaptation techniques aim to reduce the gap between two domains, which is similar to the domain generalization technique.  Moreover, the standard knowledge distillation approaches are based on a teacher-student training scheme, where a teacher model first learns the task and distills the knowledge to a student model, which is more compact with less trainable parameters. Several studies \citep{meng2018,pmlr-v139-feng21f,Xiaobo2022} attempted to train many teachers models in the source domains and ensemble them to distill knowledge into the student approach. However, because these techniques use several pretrained teacher models, they are computationally expensive to train. Recently, the knowledge distillation technique was further adopted for continual learning tasks to keep the network's responses on the previous tasks unchanged while updating it with new training samples \citep{CL_Mib}. As a result, this helped to reduce the effect of catastrophic forgetting after each round of finetuning on a new task and improve the performance.

\begin{figure}
    \centering
    \includegraphics[width=0.48\textwidth]{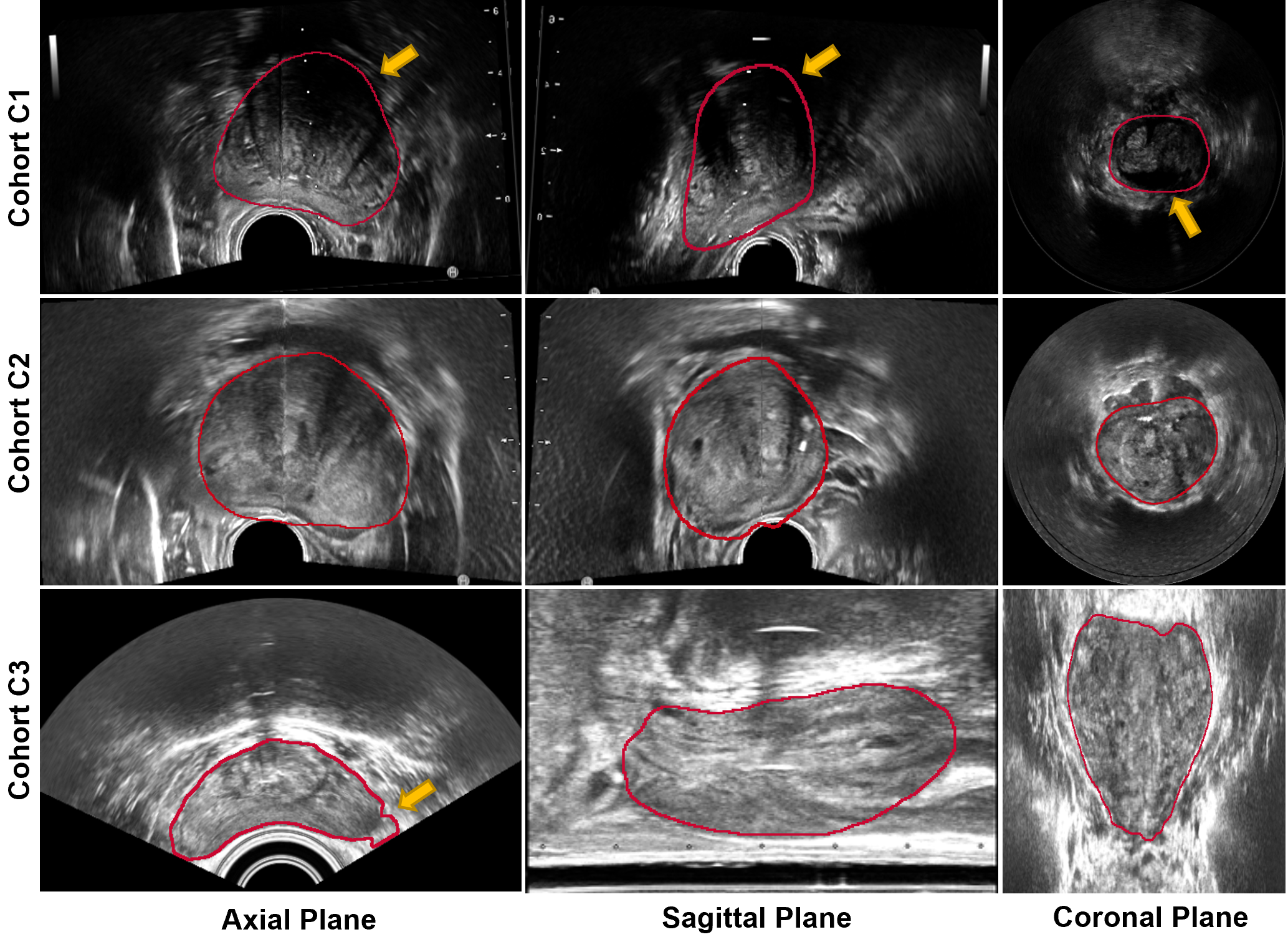}
    \caption{Example of prostate outlines (red) on ultrasound images acquired at three different institutions. Note the large variations in prostate shape, contrast, field of view, and the presence of artifacts (e.g., inhomogeneous intensity distributions). Moreover, the prostate boundary is not always clearly visible or easily distinguishable from the neighboring structures (orange arrows).}
    \label{fig:slices}
\end{figure}

In this paper, we present an end-to-end deep learning-based segmentation model for prostate gland segmentation on TRUS images. We introduce an approach for model generalization that utilizes a knowledge distillation loss \citep{lkd2019} to mitigate ``catastrophic forgetting'' while applying the model to images from multiple institutions. We first train a segmentation model on our large cohort of in-house TRUS images (cohort C1, n=764), then finetune the model on subjects from two other institutions (cohorts C2, C3) with relatively few annotated examples by transferring information from the first model to the subsequent models (\figurename~\ref{fig:slices}). 

Our study has three main contributions: 
\begin{itemize}
    \item We introduce a deep learning framework for accurate prostate gland segmentation in TRUS images, with the presence of considerable variation in intensity and image acquisition parameters. We improve the generalization capabilities of our model across data from three institutions.
    \item To limit the effect of catastrophic forgetting during transfer learning, we adapted a training scheme that utilizes knowledge distillation loss during the finetuning process on new data.
    \item Extensive experiments on multi-center data with different ultrasound probes demonstrate the proposed approach brings substantial gains over existing approaches.
\end{itemize}

\section{Methodology}
\subsection{Data and cohort description}
To develop and validate our prostate gland segmentation algorithm, our study included patients from three different institutions. This retrospective chart review study was approved by the Institutional Review Board (IRB) of Stanford University. As a chart review of previously collected data, patient consent was waived. The device and clinical characteristics of these data cohorts are listed in Table \ref{tab:addlabel}. Examples of TRUS images are shown in \figurename~\ref{fig:slices}.

\textbf{Cohort C1} included 954 men who underwent TRUS-MRI targeted biopsy utilizing the Artemis biopsy system (Eigen, Grass Valley, CA) at Stanford. Ultrasound scans were carried out using the Hitachi Hi-Vision 5500 7.5 MHz or the Noblus C41V 2-10 MHz end-fire ultrasound probe. The 3D scans were obtained by rotating the end-fire probe 200 degrees around its axis and interpolating to resample the volume to isotropic resolution. The prostate gland was segmented by an expert urologist during the TRUS-MRI targeted biopsy procedures. 

\textbf{Cohort C2} included 1,161 men who underwent biopsy at the University of California Los Angeles \citep{Natarajan2020, SONN201386, Clark2013}. Hitachi Hi-Vision 5500 7.5 MHz end-fire ultrasound probes were used for ultrasound scanning. The volume was resampled with an isotropic resolution by rotating the end-fire probe 200 degrees about its axis and interpolating. This cohort is similar to cohort C1 in terms of the ultrasound image reconstruction method.

\textbf{Cohort C3:} included TRUS scans acquired from 106 men as part of the SmartTarget Biopsy Trial \citep{Ghavami2018} who underwent targeted transperineal biopsy. For each patient, a continuous rotational 3D acquisition was used to acquire 50-120 sagittal slices to cover the whole prostate. These images were acquired using side-fire ultrasound probes. Three trained biomedical engineering experts segmented the prostate in all images where the prostate gland was visible \citep{Ghavami2018}. 

\begin{table*}[htbp]
  \centering
  \caption{Scanner specifications and demographic breakdown of the data included in our study.}
    \begin{tabular}{|llll|}
    \hline
    \textbf{Cohort} & \textbf{C1} & \textbf{C2} & \textbf{C3} \\
    \hline
    Source & In house & Public & External \\
    Number of Patients & 764 (training), 190 (testing) & 10 (training), 1,151 (testing) & 10 (training), 96 (testing) \\
    Number of Images & 802 (training), 220 (testing) & 10 (training), 1,751 (testing) & 10 (training), 96 (testing)  \\
    \hline
    Manufacturer & Eigen Inc. & Eigen Inc. & Hitachi \\
    Scanner & Artemis & Artemis & HI VISION Preirus \\
    Probe & Hitachi/Noblus/Hitachi C41V & Hitachi/Noblus/Hitachi C41V & Hitachi C41L47RP \\
    Probe Type & End-fire & End-fire & Side-fire \\
    Transmit Frequency (MHz) & 7.5 & 7.5, 10  & 3.5 \\
    Frame Rate (Hz) & 9 & 9 & 9 \\
    Average Pixel Spacing (mm) & [0.11, 0.52] & [0.21, 0.55] &  [0.25] \\
    Slice Thickness (mm) &  [0.11, 0.52] &  [0.21, 0.55] & [0.25] \\
    Matrix Size (range) & 290$\times$290 - 496$\times$496 & 342$\times$342 - 452$\times$452  & 303$\times$ 495 \\
    Number of Frames (range) &  210 - 291 &  226 - 290 & 50 - 120 \\
    \hline
    \end{tabular}%
  \label{tab:addlabel}%
\end{table*}%

\subsection{Data Preprocessing}
Multiple preprocessing steps were applied to the TRUS images. Bi-linear interpolation was used to resample the images to the same spacing ($0.25 mm \times 0.25 mm$) and to resize to $128 \times 160$ pixels, while maintaining the aspect ratio. No resampling was performed on the $z$-axis, and all slices were included in the training. For all studies in the three cohorts, the original TRUS pixel intensities (ranging between [0, 255]) were mapped to the 0-1 range using a min-max normalization. No cropping was applied as our models seek to both localize and segment the prostate. To improve the contrast of the TRUS images, the contrast limited adaptive histogram equalization \citep{Pizer1990} method was used with a default window size of $4\times4$ pixels.

\textbf{Train-test splits:} For cohort C1, 764 patients (n=802 TRUS scans) were used for training and validation of the models and 190 patients (n=220 TRUS scans) for testing. Of the 1,151 patients (n=1,761 TRUS scans) in cohort C2, ten patients (n=10 TRUS scans) were randomly chosen for finetuning the model, and the rest of the subjects were used for model testing. Similarly, for cohort C3  (n= 109 TRUS scans), only ten scans were used to finetune the model, as detailed in Table \ref{tab:addlabel}. Some patients in cohorts C1 and C2 had multiple TRUS studies.

\subsubsection{Segmentation Network Architecture}
\begin{figure*}[!t]
    \centering
    \includegraphics[width=0.95\textwidth]{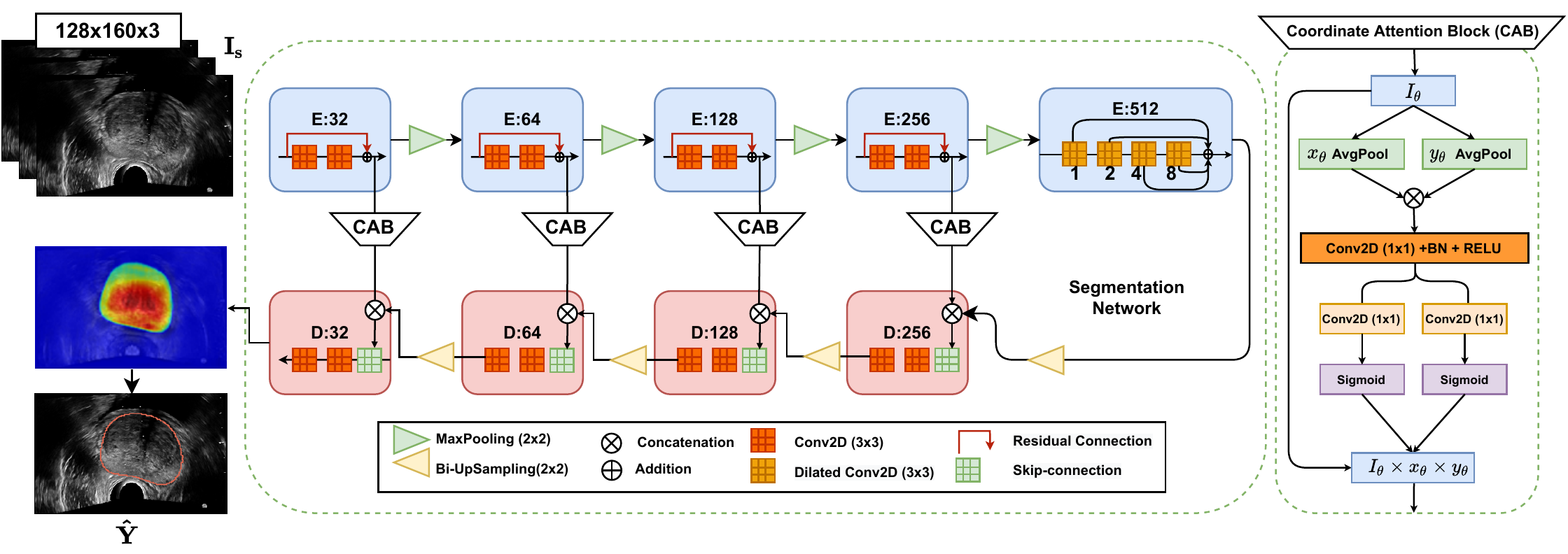}
    \caption{Flowchart of the dilated Residual UNet with coordination attention block and 2.5D input (128$\times$160 of three neighboring slices used to provide spatial context). The output of the model is the probability map of the prostate segmented in the central slice. The attention blocks assist the model in reducing uncertainty in prostate borders, which are well known to be challenging to segment due to a lack of distinct borders. $\mathbf{I}_{s}$ is the input data. $E$ and $D$ refers to encoder and decoder blocks. $\mathbf{\hat{Y}}$ is the predicted segmentation output.}
    \label{fig:DR_UNet}
\end{figure*}

Leveraging multi-scale information and boundary region guidance can help the segmentation model extract more discriminative features for robust prostate segmentation, especially given the large variation in the field of view and natural variation in prostate shapes. Thereby, we constructed a 2.5D convolutional network architecture called Coordination Dilated Residual-UNet (CoordDR-UNet) (Fig. \ref{fig:DR_UNet}) inspired by \citep{vesal2020}. The segmentation model $\mathbf{G}$ has encoder and decoder paths that are connected by a bottleneck block.  Every block in the encoder and decoder paths has two 2D convolution layers followed by a Rectified Linear Unit (ReLU), batch-normalization, and a 2D max-pooling layer to reduce the dimensionality of the feature maps. A residual connection was introduced to each encoder block to optimize the flow of gradients and force the encoder to extract more discriminative features \citep{resdeep}. A \textit{softmax} activation function was used in the model's last layer to generate the probability segmentation map of the prostate and background. The bottleneck convolution layers of UNet are replaced by dilated convolutions. This allows the model to collect both global and local contextual information by expanding the receptive field \citep{wang2018smoothed, vesal2020}. Therefore, we constructed a block of stacked dilated convolutions, the outputs of which are summed. To address the issue of gridding artifacts, each subsequent layer has complete access to earlier features learned using different dilation rates. In our model configuration, we employed four dilated convolutions in the model bottleneck with a dilation rate of $1-8$.

Moreover, our approach relies on attention mechanisms to force the deep learning model to pay more attention to the uncertain regions during model back-propagation\citep{attunet2019, 8578843, Roy_TMI2018}, specifically caused by the absence of clear boundaries between the prostate and the surrounding tissue. We attached a coordinate attention block (CAB) \citep{Hou_2021_CVPR} to our 2.5D DR-UNet to assist the model in improving the expressive power of the learned features. The coordinate attention block takes the output of each encoder block as the input $\mathcal{X_{\theta}} = [x_{1}, x_{2}, . . . , x_{N} ] \in \mathbb{R}^{h \times w \times n}$ and outputs a transformed tensor with augmented representations, i.e., feature maps,  $\mathcal{Y_\theta} = [y_{1}, y_{2}, . . . , y_{N} ]$. Here $n$ denotes the number of feature maps for each encoder block. There are two 1D global average pooling layers of $x_{\theta}$ and $y_{\theta}$ to encode each channel along with the horizontal and vertical coordinates. Each average pooling concentrates on one coordinate direction and combines information from two spatial directions, providing a pair of feature maps that are direction-aware \citep{8578843}.

\subsection{Continual Prostate Gland Segmentation}
In supervised domain adaptation or finetuning for semantic segmentation, we are given a set of source TRUS images (e.g., from Cohort C1) and their corresponding mask labels in the source domain $\mathbf{D}_{s} = {(\mathbf{I}_{s}^{i}, \mathbf{Y}_{s}^{i})}_{i=1}^{m_{s}}$, where $\mathbf{I}_{s} \in \mathbb{R} ^{w \times h \times 3}$ is the stack of three consecutive slices in the TRUS exam, $\mathbf{Y}_{s} \in \mathbb{R} ^{w \times h \times c}$ is the prostate segmentation corresponding to $\mathbf{I}_{s}$, $m_{s}$ is the number of source images and $c=2$ is the number of labels, i.e., prostate and background. In the target site, we are given few labelled images $\mathbf{D}_{t} = {(\mathbf{I}_{t}^{i})}_{i=1}^{m_{t}}$, where $m_{t}$ is the number of target images. Our goal is to train a supervised model on $\mathbf{D}_{s}$ and transfer information from $\mathbf{D}_{s}$ to reduce the gap between two domains, and improve segmentation the accuracy on $\mathbf{D}_{t}$.  

\begin{figure*}[bht]
    \centering
    \includegraphics[width=0.94\textwidth]{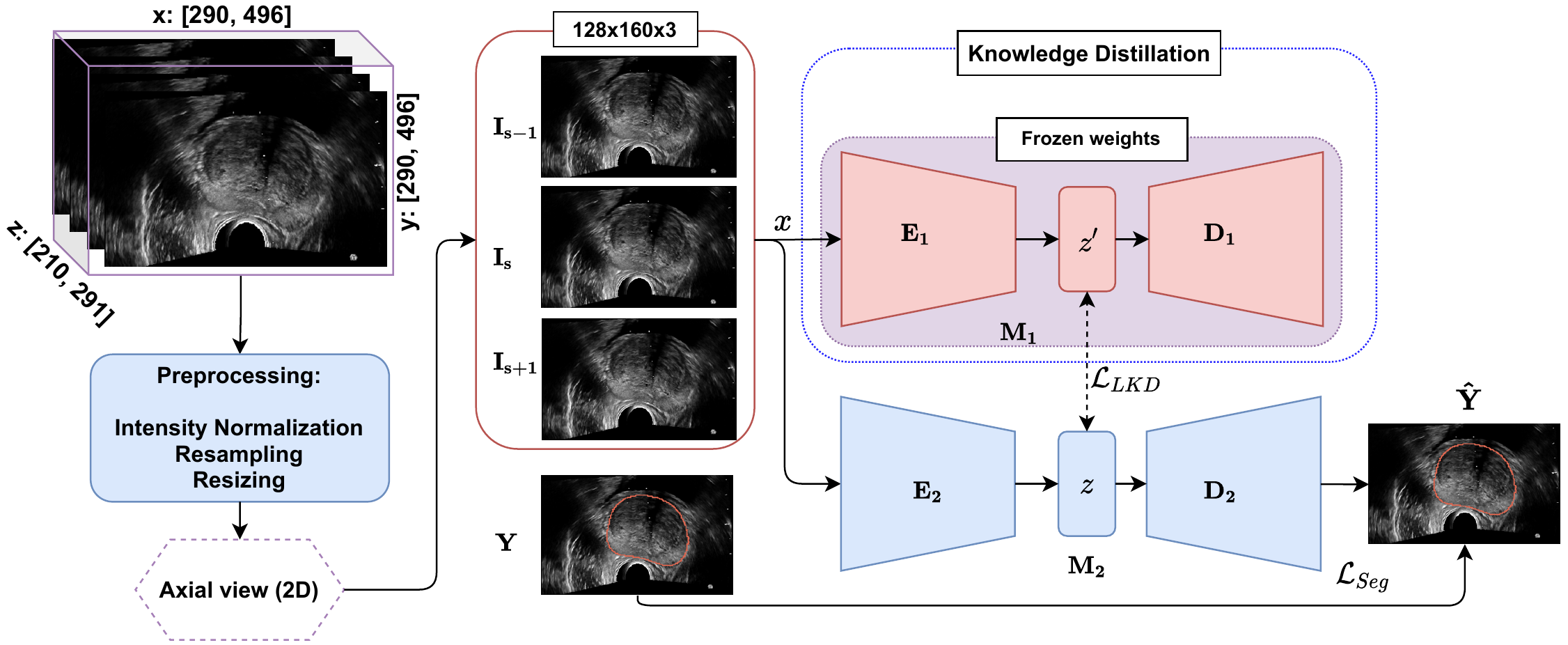}
    \caption{The overall pipeline for continual prostate gland segmentation. The segmentation model $\mathbf{M}_{s}$ is trained on cohort C1 in the first step, then the model is finetuned on cohorts C2 and C3. During optimization, the consistency of the features latent space of $\mathbf{M}_{s}$ and $\mathbf{M}_{t_{1}}$ is maintained by including the  L2 norm loss (eq. 2). The same steps are repeated for finetuning the model trained on $\mathbf{M}_{s}$ and $\mathbf{M}_{t_{1}}$ for $\mathbf{M}_{t_{2}}$  model. $\mathbf{E}_{1},\mathbf{E}_{2},\mathbf{D}_{1},\mathbf{D}_{2}$ are the encoders and decoders and $z$ and $z^{'}$  are the latent space features for the source and target models. $\mathbf{Y}$ is the ground-truth and $\mathbf{\hat{Y}}$ is the predicted segmentation output.}
    \label{fig:arch}
\end{figure*}

We were motivated by the work of \citep{lkd2019} to develop a pipeline that not only produced good segmentation accuracy for the prostate gland but also reduced the impact of model weight changes during the supervised domain adaptation process. Our approach has three steps. 
\begin{itemize}
    \item First, we trained in a supervised fashion the segmentation model $\mathbf{M}_{s}$ that segments the prostate in TRUS images using the training data from cohort C1, also referred here as  $\mathbf{D}_{s}$ (\figurename~\ref{fig:arch}). The input to the model $\mathbf{M}_{s}$ is $\mathbf{I}_{s}$, which includes a TRUS slice and two neighboring slices from the 3D TRUS volume to construct a three-channel input. The model is trained in a supervised fashion with a multi-class loss function. After training, we saved the obtained model weights as $\mathbf{M}_{s}$.
    
    \item Second, we trained a new model $\mathbf{M}_{t_{1}}$ based on $\mathbf{M}_{s}$ that takes as the input the TRUS images from cohort C2 (10 annotated cases) for finetuning. Here, the input images were fed to both models $\mathbf{M}_{t_{1}}$ and $\mathbf{M}_{s}$ while the weights for $\mathbf{M}_{s}$ were frozen. Our goal was to distill knowledge from the learned model $\mathbf{M}_{s}$ to $\mathbf{M}_{t_{2}}$ by enforcing the consistency of the latent feature space. This was achieved by minimizing the distance between $z$ and $z^{'}$ using knowledge distillation loss function as a regularization term along with a supervised segmentation loss function.
    
    \item For the third step, we repeated step two by creating a new model $\mathbf{M}_{t_{2}}$ based on trained model $\mathbf{M}_{t_{1}}$, which takes as the input the TRUS images from cohort C3 (10 annotated cases). In this step, we distill knowledge from learned model $\mathbf{M}_{t_{1}}$ to $\mathbf{M}_{t_{2}}$ similar to step two by minimizing the distance between the latent feature space. 
\end{itemize}

\subsubsection{Loss functions}
\noindent \textbf{Segmentation Loss:} The segmentation model is trained with a soft-Dice loss. To segment a TRUS image $\mathbf{I}_{s} \in \mathbb{R}^{h \times w \times 3}$ the output of \textit{Softmax} layer is two probability maps for classes $k = 0, 1$ (background and prostate) where for each pixel $\sum_{c}\mathbf{Y}_{n,k} = 1$. Given the ground-truth label $\mathbf{Y}_{n,k}$ for that identical pixel, the soft Dice loss is computed as follows:
\begin{equation}
\label{eq1lo}
	\mathcal{L}_{seg}(\mathbf{Y}, \mathbf{\hat{Y}})  = 1- \frac{1}{K}(\sum_{k}\frac{2\sum_{n}\mathbf{Y}_{nk} \mathbf{\hat{Y}}_{nk}}{\sum_{n}\mathbf{Y}_{nk} + \sum_{n}\mathbf{\hat{Y}}_{nk}})
\end{equation}

\noindent \textbf{Feature-Space Knowledge Distillation Loss:}
Knowledge distillation \citep{hinton2015} is an efficient technique for transferring knowledge from a well-trained model to a model with limited annotated data. It also has been widely used in continual learning frameworks \citep{lkd2019} when a trained model is updated on a new task that interferes with the learned representations on the previous task. We hypothesize that the same catastrophic forgetting problem exists for domain generalization across multi-institutional data. The $\mathbf{M}_{s}$ model (global model) is updated on a small set of randomly sampled data for each cohort at each round, while the cohorts have different distributions from the previous round. Therefore, to reduce the impact of model weight changes in the latent feature space of $\mathbf{M}_{t_{1}}$ and $\mathbf{M}_{t_{2}}$ during finetuning, an L2-norm was applied as a knowledge distillation loss. This regularization loss enforced the model to preserve the previous knowledge and keep the latent features space of $\mathbf{M}_{t_{n}}$ the same, where n is the number of available cohorts.  

\begin{equation}
    \centering
    \mathcal{L}_{KDL}  = \frac{||\mathbf{E}_{1}(z) - \mathbf{E}_{2}(z^{'} )||_{2}^{2}}{|\mathbf{D}_{t}|}
\end{equation}

where $z$ and $z^{'}$ denotes the latent features computed by model encoder $\mathbf{E}_{1}$ and $\mathbf{E}_{2}$ when a TRUS image from cohort C2 or C3 is fed.

The overall loss function for training the whole pipeline is defined as:
\begin{equation}
    \mathcal{L}_S = \mathcal{L}_{seg} + \lambda\mathcal{L}_{KDL}
\end{equation}
where $\lambda$ is the weight that controls the impact of knowledge distillation term during optimization.

\subsection{Implementation details}
The prostate gland segmentation framework was trained in an end-to-end fashion using the Adam optimizer with an initial learning rate of $\eta = 10^{-3}$ and exponential weight decay $\alpha =0.01 $ for 500 epochs, with a batch size of 64. The momentum parameters for the Adam optimizer were set to $\beta_{1} = 0.9$ and $\beta_{2}=0.999$. The learning rate was reduced on the plateau if the validation loss did not improve after every 30 epochs. We used early stopping if no reduction in validation loss was seen for 50 epochs. During training, we utilized an online data augmentation library (imgAug~\cite{imgaug}) and employed a set of random data augmentation for each input image $\mathbf{I}_{s} \in \mathbb{R}^{h \times w \times 3}$. The augmentation includes horizontal/vertical axis flipping, image scaling with a factor of $[0.8, 1.2]$, Gaussian noise with $\sigma \in [0.0 , 0.3]$, elastic deformation with the displacement field strength of $\alpha \in [0.5, 0.35]$ and a displacement field smoothness of $\sigma = 0.25$. To evaluate our method, we compared the resulting segmentations with the ground-truth masks using several metrics namely Dice similarity coefficient (Dice), Hausdorff Distances (HD), and sensitivity. We used the 95th percentile of the Hausdorff distances (HD95) between model prediction $\mathbf{Y}$ and ground-truth mask $\mathbf{\hat{Y}}$ to eliminate the impact of outliers. The paired Student’s t-test was used to assess the statistical significance of Dice differences when comparing multiple methods \citep{Nadeau2003}. 

We used PyTorch 1.7 \citep{NEURIPS2019_9015} to develop and train all models. Training took around 24 hours on an NVIDIA RTX A6000 GPU with 32GB memory. The code is available online at: \hyperlink{https://github.com/pimed/TRUSGlandSegmentation}{https://github.com/pimed/TRUSGlandSegmentation}

\section{Results}
Several experiments were carried out to 1) determine the best segmentation model on cohort C1-test and 2) identify the best strategy for knowledge transfer during finetuning on data from other institutions. 

\subsection{Prostate Gland Segmentation}
\begin{figure*}[ht]
    \centering
    \includegraphics[width=\textwidth]{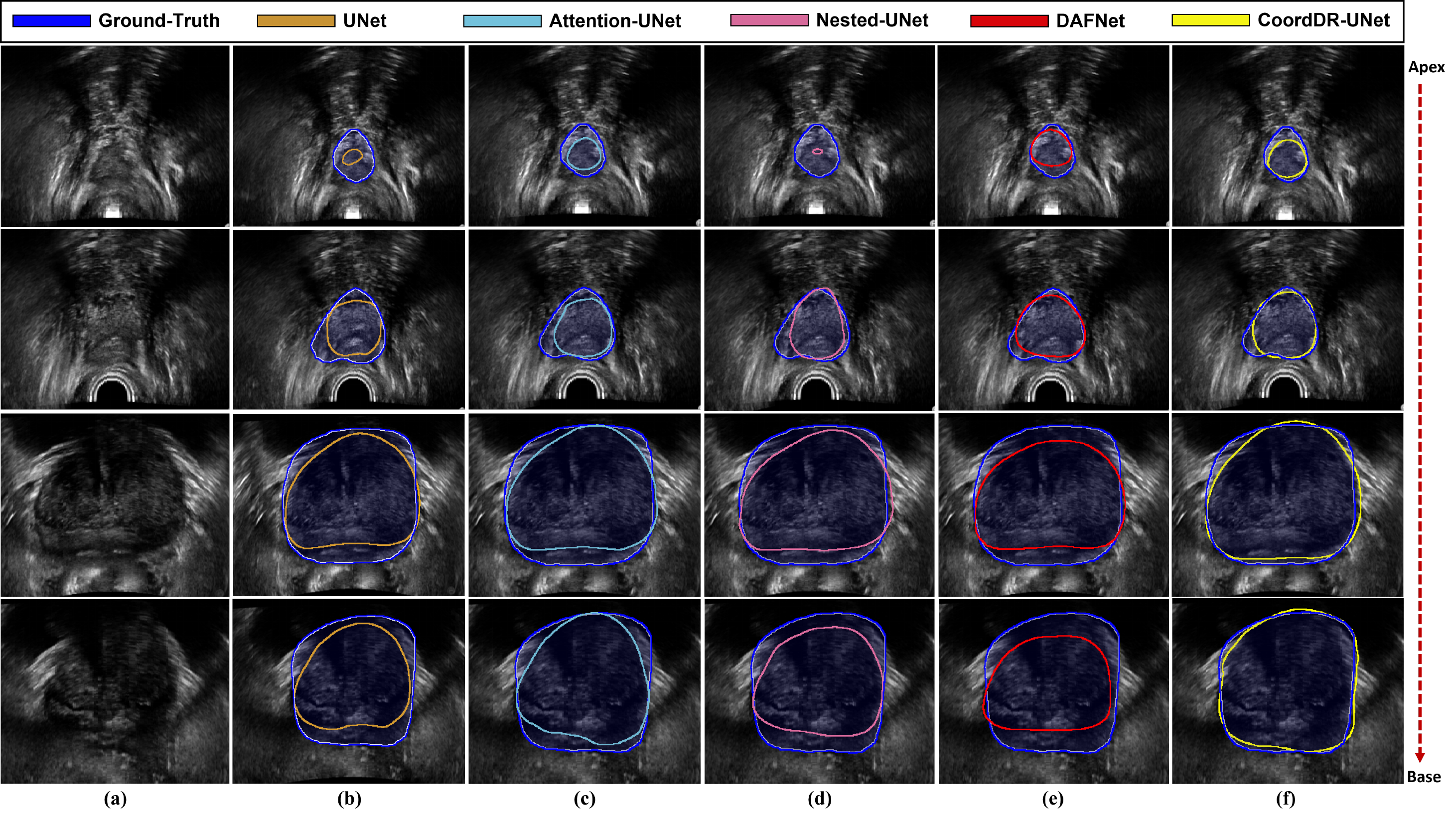}
    \caption{Visual comparison of prostate segmentation results produced by different methods for a sample patient in cohort C1-test. From left to right are the input TRUS slices (column (a)), the UNet 2.5D predictions (column (b)),  Attention-UNet 2.5D (column c), Nested-UNet 2.5D (column (d)), DAFNet \citep{wang2019} (column (e)) and our proposed CoordDR-UNet 2.5D (column (f) for different slices from the apex (top row) to base (bottom row). The blue contours show the ground-truth segmentation outlined by an expert urologist.}
    \label{lab:C1_seg}
\end{figure*}

We compared CoordDR-UNet with several well-known segmentation approaches, including UNet, Attention-UNet, Nested-UNet, Dilated-residual UNet, and Deep Attentive Features Network (DAFNet)  \citep{wang2019} which is a new approach for 3D segmentation of TRUS. All five approaches were trained using their public implementations, and their training parameters were adjusted to obtain the best segmentation results. Fig. \ref{lab:C1_seg} shows four slices of a test set patient in cohort C1 from apex to base (column (a)). It also shows comparative performance among the UNet (column (b)), Attention-UNet (column (c)), Nested-UNet (column (d)), DAFNet (column (e)) and CoordDR-UNet (column (f)). CoordDR-UNet segmented the prostate successfully, with better prediction at the apex and the base of the prostate (rows 1,4). Furthermore, for mid-gland slices (row 2-3), CoordDR-UNet produced very accurate prostate gland mask predictions compared to other state-of-the-art methods. 

Table \ref{tab:UCIL_tab} summarizes the results for cohort C1-test for all models using several segmentation metrics. To speed up volumetric segmentation while minimizing memory requirements and incorporating contextual and temporal information for the model, all models were trained with a 2.5D input. As seen in Table \ref{tab:UCIL_tab}, the 2.5D UNet outperformed the 2D and 3D UNet with a Dice score of $0.91\pm0.07$. The surface distance metric, HD95,  was also reduced, i.e., 3.0 $mm$ for 2.5D UNet vs 4.03 $mm$ for 2D UNet. The performance of Attention-UNet and Nested-UNet was similar to that of UNet. DAFNet, which is a recent model specifically designed for prostate gland segmentation in 3D TRUS achieved a Dice score of $0.92\pm0.03$ and HD95 value of 2.87 $mm$. Our proposed CoordDR-UNet approach, which incorporates coordinate attention blocks and dilated convolutions, outperformed other models, with a Dice score of $0.94\pm0.03$ and the lowest HD95 score of 2.29 $mm$ compared with the other approaches. We will refer to this model as the model $\mathbf{M}_s$. Figure S2 in the supplementary material shows the impact of training data size with respect to segmentation accuracy.

\begin{table}[ht!]
    \caption{Quantitative comparison results $(mean (\pm std))$ between the proposed method and other segmentation methods for cohort C1-test dataset. The best results are highlighted in bold. Paired Student’s t-tests showed statistical significance (P $\leq$ 0.05) when comparing the methods with the baseline method 2D UNet. DAFNet: Deep Attentive Features Network, DR-UNet: Dilated-Residual UNet, CoordDR-UNet: Coordination Dilated-Residual UNet, HD95: 95th percentile of the Hausdorff distances.}
    
    \label{tab:UCIL_tab}
    \centering
    \resizebox{0.48\textwidth}{!}{\begin{tabular}{@{}l|c|lll@{}}
    \toprule
                      & \multicolumn{4}{c}{\textbf{Evaluation on cohort C1-test (n=220)}}   \\ \cmidrule(r){1-5}
    Methods & \multicolumn{1}{l|}{Input} &  Dice   & HD95 $[mm] $ & Sensitivity       \\ \midrule
    UNet \scalebox{0.8}{\citep{Unet2015}} & \multirow{2}{*}{2D}   & 0.90 ($\pm$0.03) & 4.03 ($\pm$3.52)  & 0.94 ($\pm$0.04) \\ \cmidrule(r){3-5}
    CoordDR-UNet (proposed) \scalebox{0.8} &   & 0.92 ($\pm$0.03) & 3.29 ($\pm$1.45) & 0.92 ($\pm$0.05) \\ \cmidrule(r){1-5}
    UNet \scalebox{0.8}{\citep{Unet2015}}     & \multirow{3}{*}{3D}    & 0.90 ($\pm$0.03) & 5.21 ($\pm$3.52)   & 0.92 ($\pm$0.01)  \\ \cmidrule(r){3-5}
    DAFNet \scalebox{0.8}{\citep{wang2019}}  &   & 0.92 ($\pm$0.03) & 2.87 ($\pm$1.27)   & 0.90 ($\pm$0.04)  \\ \cmidrule(r){3-5}
    CoordDR-UNet (proposed) \scalebox{0.8}  &    & 0.92 ($\pm$0.04) & 3.00 ($\pm$1.46) & 0.92 ($\pm$0.07)  \\ \cmidrule(r){1-5}
        
    \begin{tabular}[c]{@{}l@{}}UNet \scalebox{0.8}{\citep{Unet2015}}\end{tabular} & \multirow{5}{*}{2.5D}  & 0.91 ($\pm$0.07) & 3.00 ($\pm$1.30)  & 0.91 ($\pm$0.08)  \\
    Att-UNet \scalebox{0.8}{\citep{attunet2019}} &  & 0.92 ($\pm$0.02)  & 2.75 ($\pm$1.16)  & 0.93 ($\pm$0.04) \\
    Nested-Unet \scalebox{0.8}{\citep{nestedUnet2018}}  & & 0.91 ($\pm$0.04)  & 3.44 ($\pm$1.55)  & 0.93 ($\pm$0.06)  \\
    DR-UNet \scalebox{0.8}{\citep{vesal2020}} & & 0.93 ($\pm$0.02) & 2.34 ($\pm$0.92)  & 0.94 ($\pm$0.03)  \\
    CoordDR-UNet (proposed) &  & \textbf{0.94 ($\pm$0.03)} & \textbf{2.29 ($\pm$1.45)} & \textbf{0.94 ($\pm$0.05)}  \\ \cmidrule(r){1-5} 
    \end{tabular}}
\end{table}

\begin{table}[!t]
\centering
    \caption{Quantitative comparison results $(mean (\pm std))$ for different finetuning strategies shown for cohort C2-test and C3-test. The best results are highlighted in bold. Paired Student’s t-tests showed statistical significance (P $\leq$ 0.05) when comparing the CoordDR-UNet + KDL with no pretraining and direct prediction models. Bolded entries represent the best metric in each test set. }
    \label{tab:resultsc2c3}
    \resizebox{0.48\textwidth}{!}{\begin{tabular}{@{}lllll@{}}
    \toprule
    Methods &  & Dice  & HD95 $[mm]$ & Sensitivity \\\midrule
    
                      & \multicolumn{4}{c}{\textbf{Evaluation on cohort C2-test (n=1,751)}}   \\ \cmidrule(r){2-5}
    CoordDR-UNet        &exp1& 0.89 ($\pm$0.03) & 4.03 ($\pm$1.62)  & 0.84 ($\pm$0.06)    \\ \midrule
    CoordDR-UNet + w/o pretrained &exp2& 0.71 ($\pm$0.09) & 14.8 ($\pm$4.60)  & 0.77 ($\pm$0.12)    \\ \midrule
    CoordDR-UNet + Finetuning &exp3& 0.90 ($\pm$0.03) & 3.80 ($\pm$1.60)  & 0.87 ($\pm$0.05)    \\ \midrule
    CoordDR-UNet + KDL &exp4& \textbf{0.91 ($\pm$0.03)} & \textbf{3.69 ($\pm$1.49)}  & \textbf{0.88 ($\pm$0.05)}    \\ \bottomrule
                      & \multicolumn{4}{c}{\textbf{Evaluation on cohort C3-test (n=96)}}  \\ \cmidrule(r){2-5}
    CoordDR-UNet      &exp1& 0.24 ($\pm$0.29) & 14.14 ($\pm$11.83)  & 0.18 ($\pm$0.25)   \\ \midrule
    CoordDR-UNet + w/o pretrained &exp2& 0.78 ($\pm$0.14)  & 10.23 ($\pm$6.59)  & 0.76 ($\pm$0.18)    \\ \midrule
    CoordDR-UNet + Finetuning &exp3& 0.80 ($\pm$0.19)  & 7.30 ($\pm$6.21)  & 0.73 ($\pm$0.10)    \\ \midrule
    CoordDR-UNet + KDL  &exp4& \textbf{0.82 ($\pm$0.16) } & \textbf{7.13 ($\pm$6.25)}   & \textbf{0.76 ($\pm$0.19)}   \\ \bottomrule
    
    \end{tabular}}
\end{table}

\begin{figure}[ht]
    \centering
    \includegraphics[width=0.5\textwidth]{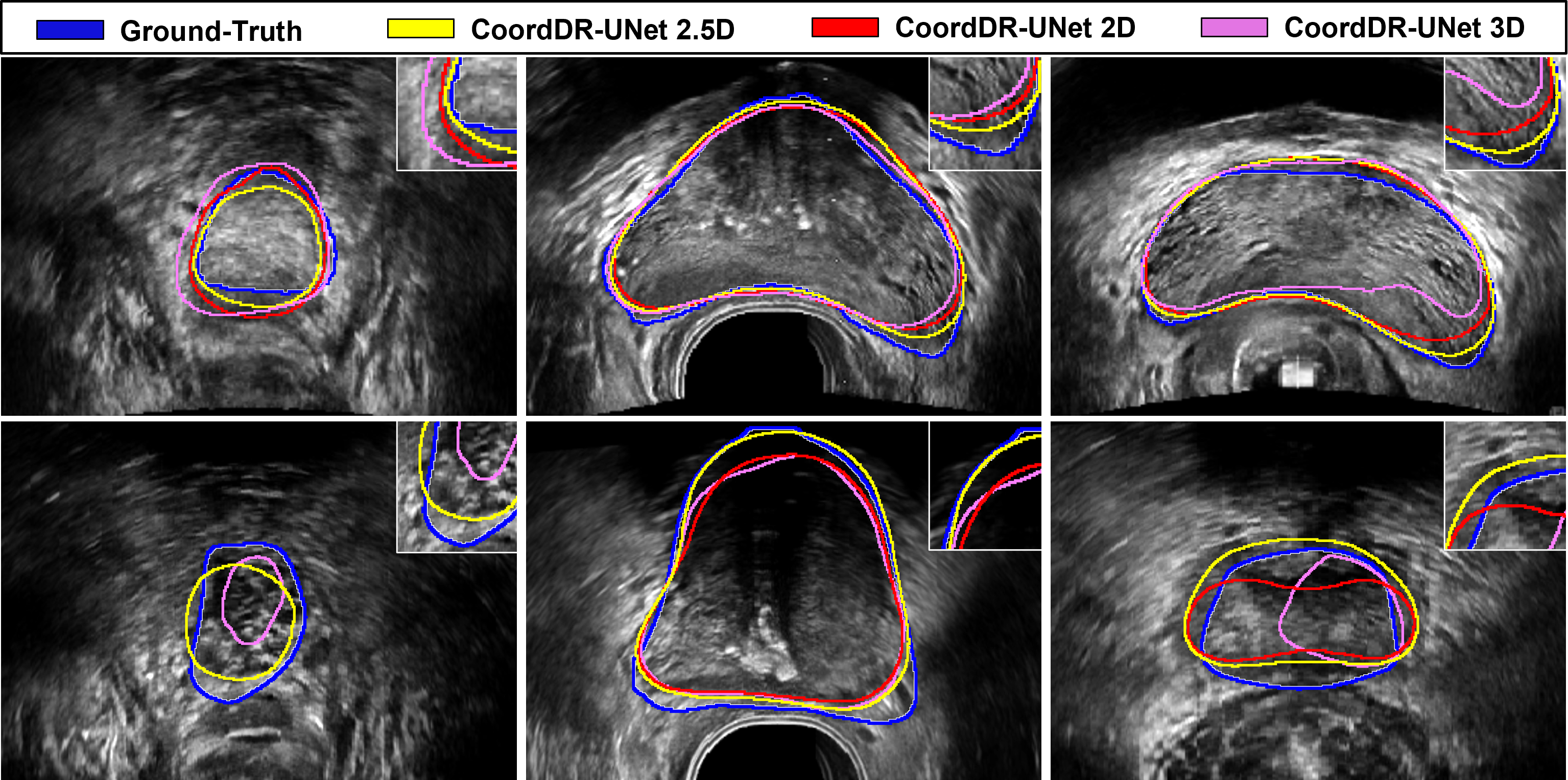}
    \caption{Visual comparison of CoordDR-UNet segmentation output with 2D, 3D, and 2.5D as the input. Each row shows apex, mid-gland, and base slices from different subjects from the cohort C1-test dataset.}
    \label{fig:2D3D}
\end{figure}

\begin{figure}[ht]
    \centering
    \includegraphics[width=0.5\textwidth]{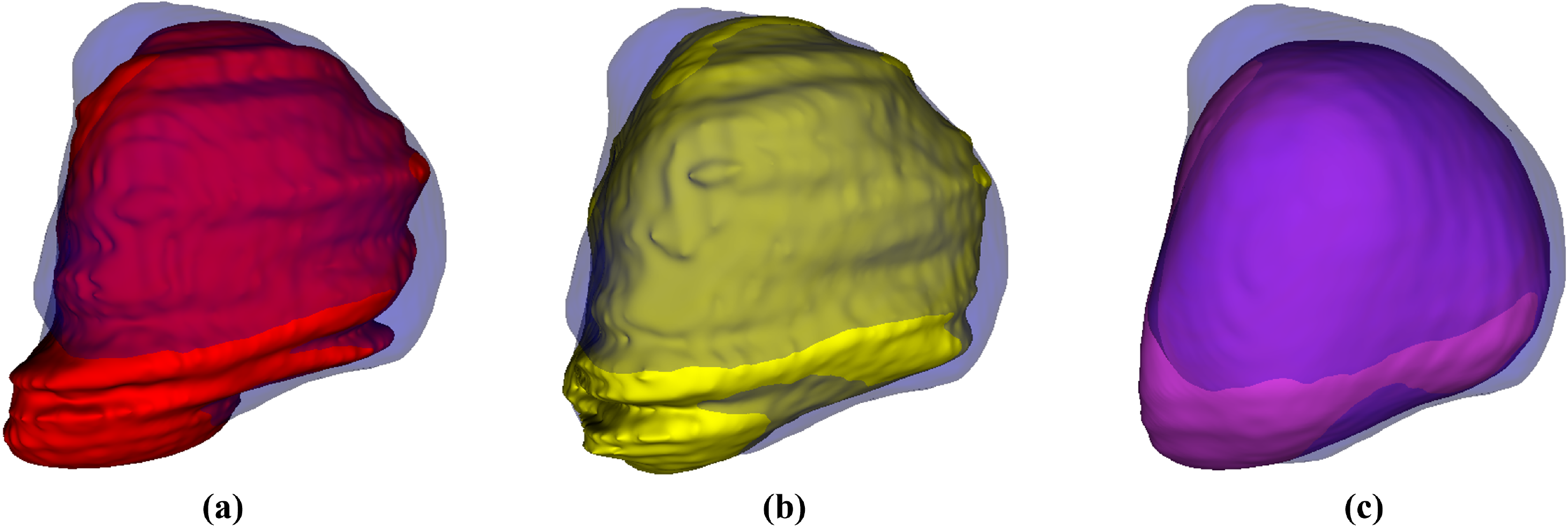}
    \caption{3D visualization of the segmentation results in a TRUS volume. CoordDR-UNet output with (a) 2D, (b) 2.5D and (c) 3D representations. The ground-truth mask is shown on a blue surface. }
    \label{fig:2D3D_difference}
\end{figure}

To show the benefit of the 2.5D input representations, we also trained the CoordDR-UNet model with 2D and 3D data representations as input. For the 3D model, the standard architecture of CoordDR-UNet was changed by replacing all 2D operations with 3D operations, including convolution, batch normalization, max-pooling, and upsampling layers. Moreover, the 3D TRUS images were downsampled to a fixed size ($80\times 160 \times 128$) similar to \citep{wang2019}. The quantitative results (Table 2) show that CoordDR-UNet 2.5D outperformed the models trained with 2D and 3D input representations, suggesting the benefit of the 2.5D representation. CoordDR-UNet with 2D input achieved a Dice score of $0.92\pm0.03$ and HD95 value of 3.29 $mm$ while CoordDR-UNet with 3D input obtained a Dice score of $0.92\pm0.04$ and HD95 value of 3.00 $mm$. Furthermore, Fig. \ref{fig:2D3D} shows visual comparison between CoordDR-UNet models trained with 2D, 2.5D and 3D input representations. One limitation of the 2D segmentation approaches for prostate gland segmentation is inconsistency across adjacent slices. The difference between CoordDR-UNet 2.5D and CoordDR-UNet with 2D and 3D input representations is shown in Fig. \ref{fig:2D3D_difference}. The 2.5D representation achieved a more accurate segmentation and reduced inconsistencies across adjacent slices. The 3D segmentation approach has smoother boundaries, but the output is less accurate when compared to the ground-truth masks. These results highlight the benefit of 2.5D  representations, which not only produced highly accurate prostate gland segmentation but also consistent segmentation across adjacent slices.

\begin{figure*}[ht]
    \centering
    \includegraphics[width=\textwidth]{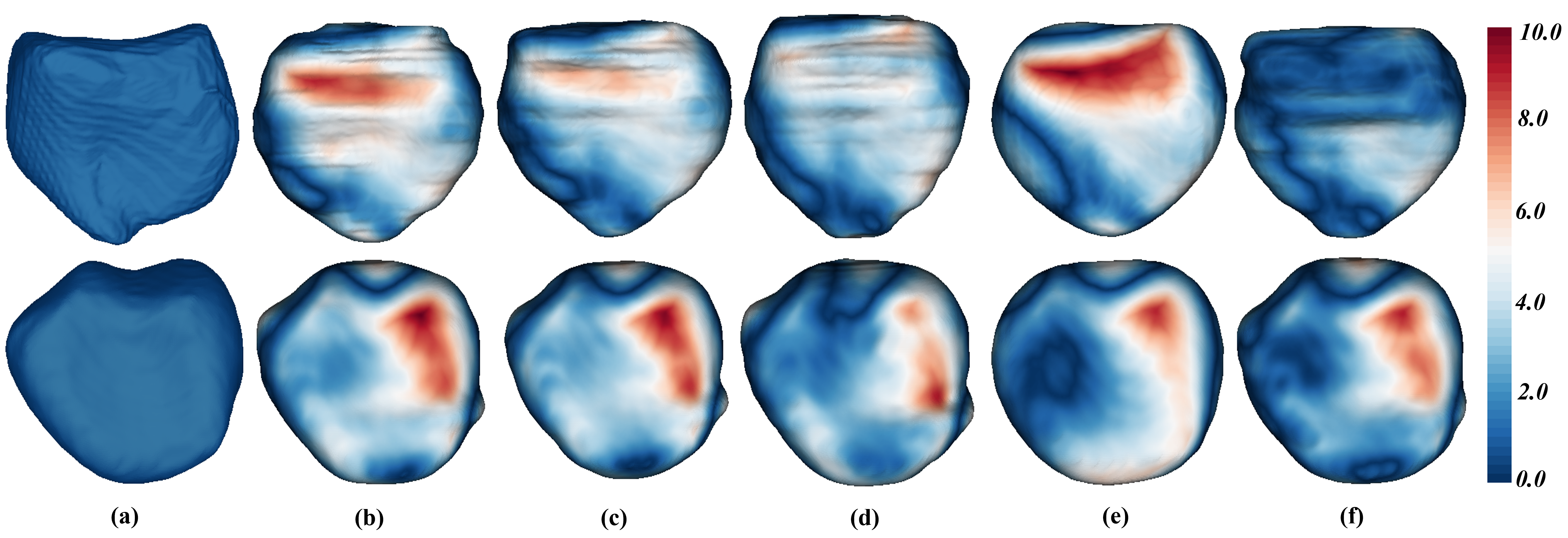}
    \caption{3D visualization of the surface distance (in voxel) between segmented surface and ground truth. Each row shows one subject. Different colors represent different surface distances. From left to right are (a) ground-truth, (b) UNet 2.5D, (c) Attention-UNet 2.5D, (d) Nested-UNet 2.5D, (e) DAFNet3D \citep{wang2019} and (f) our proposed CoordDR-UNet 2.5D. Our method consistently performs well on the whole prostate surface.}
    \label{fig:surfaceDistance}
\end{figure*}

For the error analysis, we computed the surface error between model predictions and their corresponding ground truth. Fig. \ref{fig:surfaceDistance} shows the 3D visualization of the surface distance between the ground-truth and segmentation output by different methods for two test cases from the cohort C1-test. Our method consistently achieved accurate and robust segmentation covering the whole prostate, including challenging regions such as the apex and base. CoordDR-UNet 2.5D had the lowest average HD95 surface distance (2.29 $mm$) compared to other approaches for cohort C1-test data. DAFNet produced a more smooth segmentation output because it was trained with 3D input representations. However, it had a higher surface distance error (2.87 $mm$) because the segmentation output is less accurate in comparison to ground truth.

\subsection{Model performance on independent data}
\begin{figure*}[ht]
    \centering
    \includegraphics[width=0.95\textwidth]{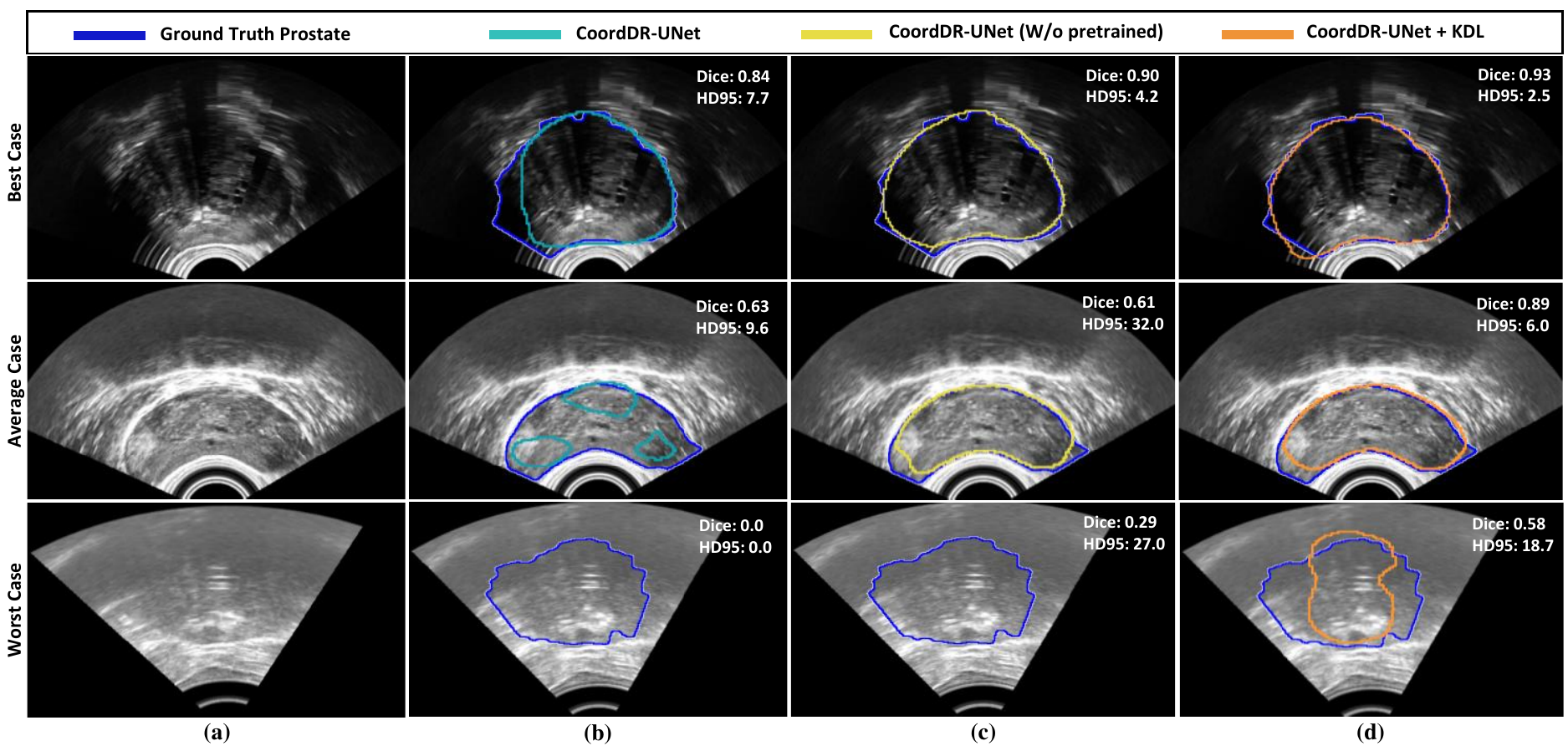}
    \caption{Visual comparison of segmentation results produced by different methods for three patients (best, average, worst) in cohort C3-test. From left to right are the 2D TRUS slices (column (a)), CoordDR-UNet 2.5D direct prediction without finetuning (column (b)), CoordDR-UNet 2.5D trained on ten TRUS images without any pretraining (column (c)), and our proposed CoordDR-UNet 2.5D finetuned on ten TRUS images with knowledge distillation loss (column (d)). Dice and HD95 scores are also shown for each patient.}
    \label{lab:C3_seg}
\end{figure*}

To assess model generalizability, we evaluated our framework on TRUS images from cohorts C2 and C3 (Table \ref{tab:resultsc2c3}). We considered four different scenarios to demonstrate the performance of CoordDR-UNet 2.5D with knowledge distillation. In the first scenario, we directly tested the model $\mathbf{M}_1$ (trained with data from Cohort C1) on cohorts C2 and C3 without domain adaptation or finetuning. The model achieved good results on cohort C2  (Table \ref{tab:resultsc2c3}, 3rd row) with a Dice score of $0.89\pm0.03$ (P$\leq$0.05) and HD95 of 4.03 $mm$. This relatively high performance is due to the similar data acquisition and vendor for the TRUS images in cohorts C1 and C2. However, the performance of our model $\mathbf{M}_1$ on cohort C3 was poor (Dice score = $0.24\pm0.29$ (P$\leq$0.05)). This reduced performance is likely caused by the differences in acquisition, field of view, ultrasound manufacturer, and ultrasound probes (side-fire) between cohorts C1 and C2 versus C3 (Fig. \ref{lab:C3_seg}). 

In the second scenario, we obtained ultrasound scans and ground truth from the prostate segmentation from ten random subjects in each cohort (C2 and C3). The CoordDR-UNet model was trained from scratch (without pretraining) to investigate how the model performs with limited annotated data. On cohort C2-test data, this model (Table \ref{tab:resultsc2c3}, 4th row) obtained a Dice score of $0.71\pm0.09$ and HD95 value of 14.8 $mm$, which demonstrates a substantial performance drop in comparison to the CoordDR-UNet trained on cohort C1 data. This is because a model with only ten training samples cannot capture all of the variations in the test data since prostate volumes and image quality vary significantly between patients. We observed similar results when we utilized only ten examples from cohort C3 to train CoordDR-UNet from scratch. Furthermore, the model without pretraining (Table \ref{tab:resultsc2c3}, 8th row) achieved a Dice score of $0.78\pm0.14$ (P$\leq$0.05) and surface distance of 10.23 $mm$ (HD95) on cohort C3-test. Training with only ten TRUS scans from cohort C3 assisted the model to somewhat capture the overall prostate shape and geometry compared to only using the model $\mathbf{M}_s$ which had not seen any data from Cohort C3.

In the third scenario, we performed standard finetuning on CoordDR-UNet with a pretrained model and without distillation loss. All the layers in CoordDR-UNet were frozen except the last layer \citep{Weiss2016}. On cohort C2-test data, this model obtained a Dice score of $0.90\pm0.03$ (P$\leq$0.05) and HD95 value of 3.80 $mm$ (Table \ref{tab:resultsc2c3}, 5th row). On cohort C3-test data, this model achieved a Dice score of $0.80\pm0.03$ (P$\leq$0.05) and HD95 value of 7.30 $mm$ (Table \ref{tab:resultsc2c3}, 10th row). 

\begin{figure*}[ht]
    \centering
    \subfloat[Cohort C2-test]{{\includegraphics[width=0.47\textwidth]{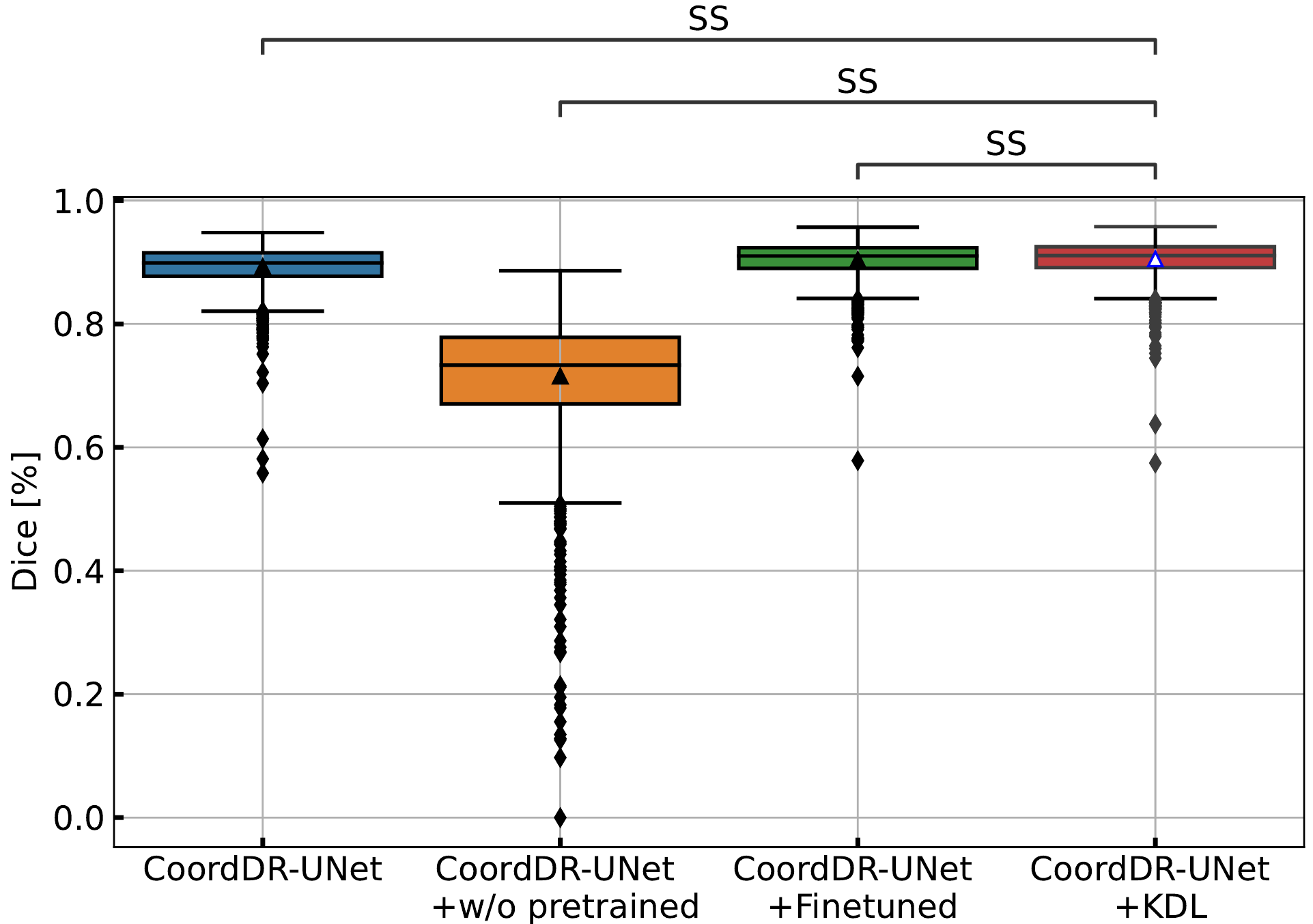} }}
    \qquad
    \subfloat[Cohort C3-test]{{\includegraphics[width=0.47\textwidth]{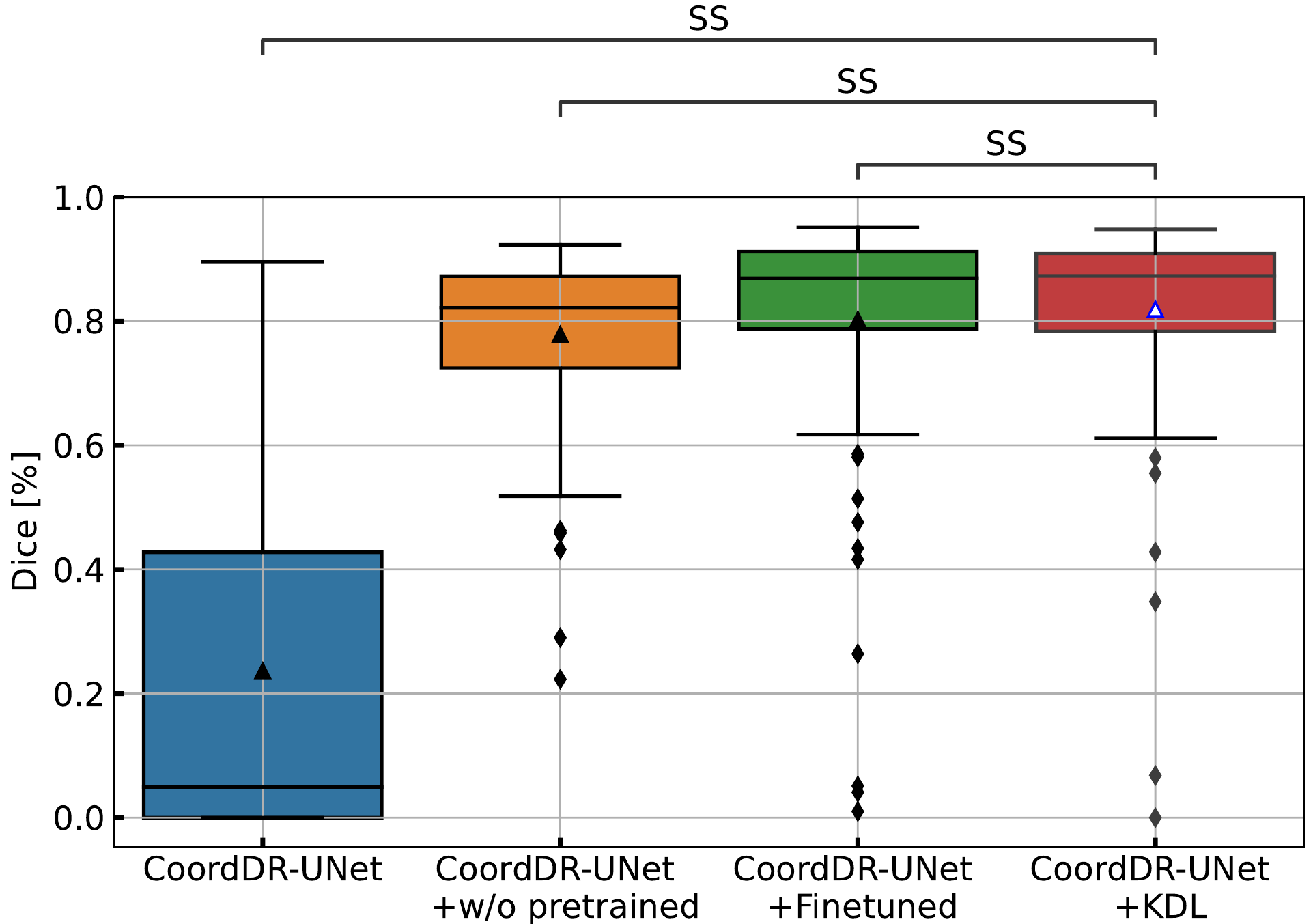} }}
    \caption{Box plots of Dice measure for the CoordDR-UNet, CoordDR-UNet + W/o pretrained, CoordDR-UNet with standard finetuning, and CoordDR-UNet + KDL segmentation approaches of cohort C2-test and C3-test. SS: statistically significant (P $\leq$ 0.05), NS: not significant (P > 0.05).}
    \label{fig:boxstat}
\end{figure*}

In the fourth scenario, we tested CoordDR-UNet with a pretrained model and knowledge distillation loss (Table \ref{tab:resultsc2c3}). In the studies from cohort C2-test, the model obtained a Dice score of $0.91\pm0.03$ and HD95 value of 3.69 $mm$ (Table \ref{tab:resultsc2c3}, 6th row), showing similar results with studies on cohort C1-test data. We refer to this model as model $\mathbf{M}_{t_1}$. Moreover, CoordDR-UNet + KDL improved the prostate gland segmentation for studies in cohort C3-test, achieving a Dice score of $0.82\pm0.16$ (P$\leq$0.05), and HD95 value of 7.13 $mm$. We refer to this model as model $\mathbf{M}_{t_2}$. The quantitative evaluation showed that the CoordDR-UNet + KDL approach achieved a significantly higher (P$\leq$0.05) Dice score compared to the CoordDR-UNet without pretraining and the CoordDR-UNet direct prediction for both cohorts (Fig. \ref{fig:boxstat}). 

\begin{figure*}[ht]
    \centering
    \includegraphics[width=\textwidth]{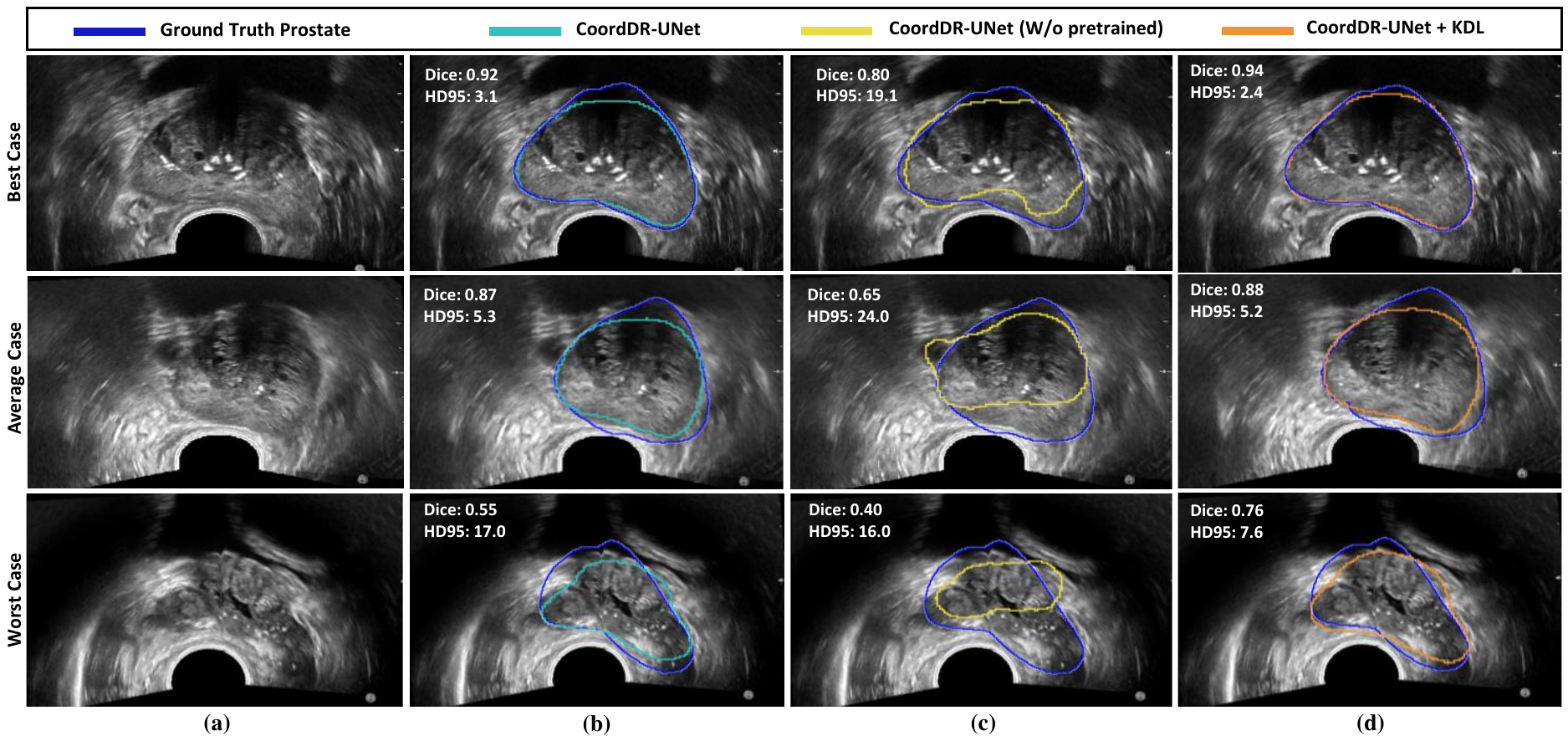}
    \caption{Visual comparison of segmentation results produced by different methods for three patients (best, average, worst) in cohort C2-test. From left to right are the 2D TRUS slices (column (a)), CoordDR-UNet 2.5D prediction without finetuning (column (b)), CoordDR-UNet 2.5D trained on ten TRUS images without any pretraining (column (c)), and our proposed CoordDR-UNet 2.5D finetuned on ten TRUS images with knowledge distillation loss (column (d)). Dice and HD95 scores are shown for each patient.}
    \label{lab:C2_seg}
\end{figure*}

\begin{figure}[ht]
    \centering
    \includegraphics[width=0.47\textwidth]{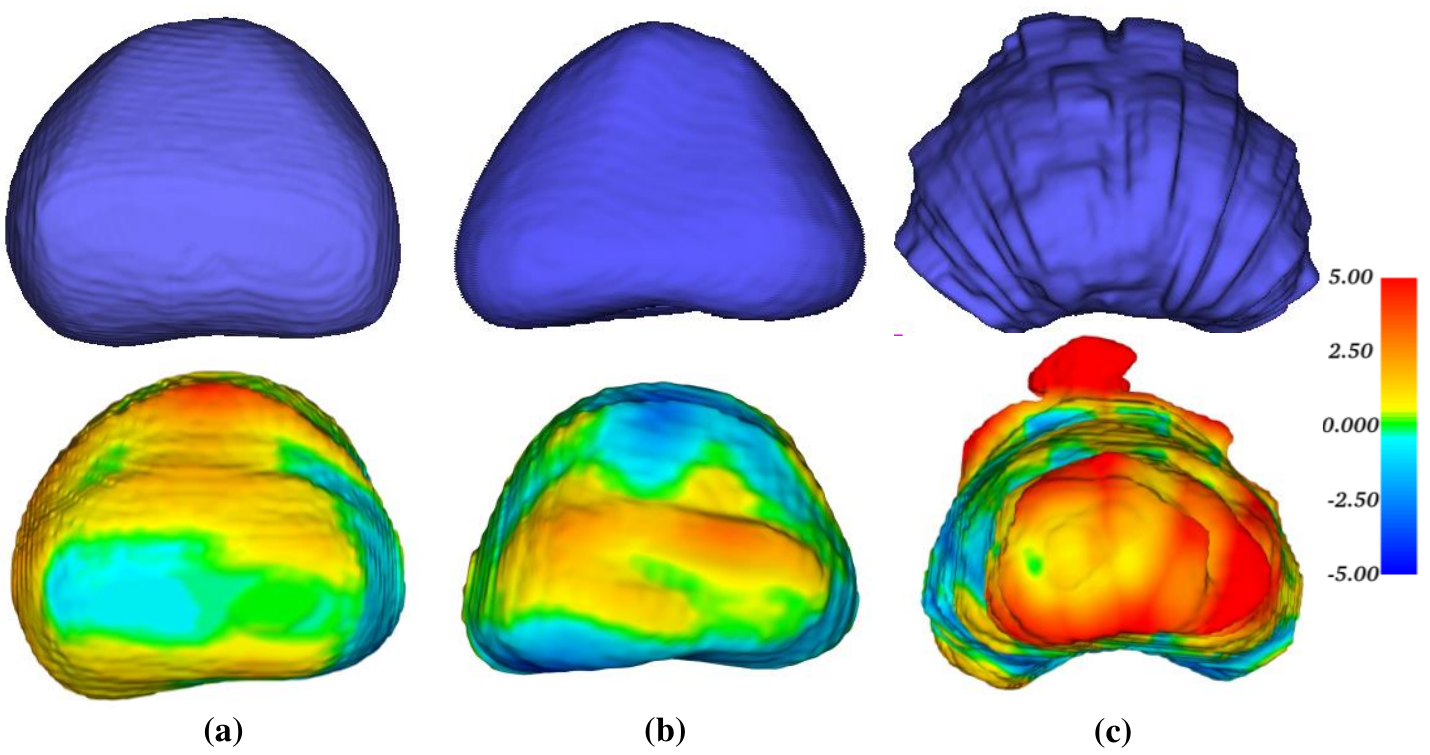}
    \caption{3D visualization of the surface distance (in voxel) between segmented prostate and ground truth. Different colors represent different surface distances. The top row shows the ground-truth surface masks for three different patients from cohort C1-test (a), C2-test (c), and C3-Test (b). The bottom row shows the computed HD surface distance between CoordDR-UNet 2.5D predictions and ground-truth.}
    \label{fig:surfacedistance}
\end{figure}

Figs. \ref{lab:C2_seg}-\ref{lab:C3_seg} display segmentation outputs for the best, average, and worst performing cases in cohorts C2-test and C3-test using the different generalization approaches we considered. In data from both cohorts, CoordDR-UNet + KDL obtained the most accurate segmentation among different training strategies. Fig. \ref{fig:surfacedistance} shows 3D segmentation results for CoordDR-UNet on TRUS images from the three cohorts, as well as the corresponding surface distance between segmented surfaces and ground truth volumes. These results highlight that the proposed method obtained accurate and smooth segmentation surfaces covering the whole prostate region for cohorts C1-C2 (columns a-b), but the surface distances are high for the test cases in cohort C3 (column c).

\subsection{Model performance after finetuning}
To evaluate the performance of our method after all iterations of finetuning on cohorts C1, C2, and C3. Table S1 in the Supplementary material shows the quantitative results for the following three models:

\begin{itemize}
    \item $\mathbf{M}_{s}$: model trained using the training data from cohort C1 
    \item $\mathbf{M}_{t_{1}}$: model $\mathbf{M}_{s}$, finetuned using 10 TRUS images from cohort C2 using CoorDR-UNet and Knowledge Distillation Loss
    \item $\mathbf{M}_{t_{2}}$: model $\mathbf{M}_{t_{1}}$, finetuned using 10 TRUS images from cohort C3 using CoorDR-UNet and Knowledge Distillation Loss
\end{itemize}

\begin{figure}[!t]
    \centering
    \includegraphics[width=0.35\textwidth]{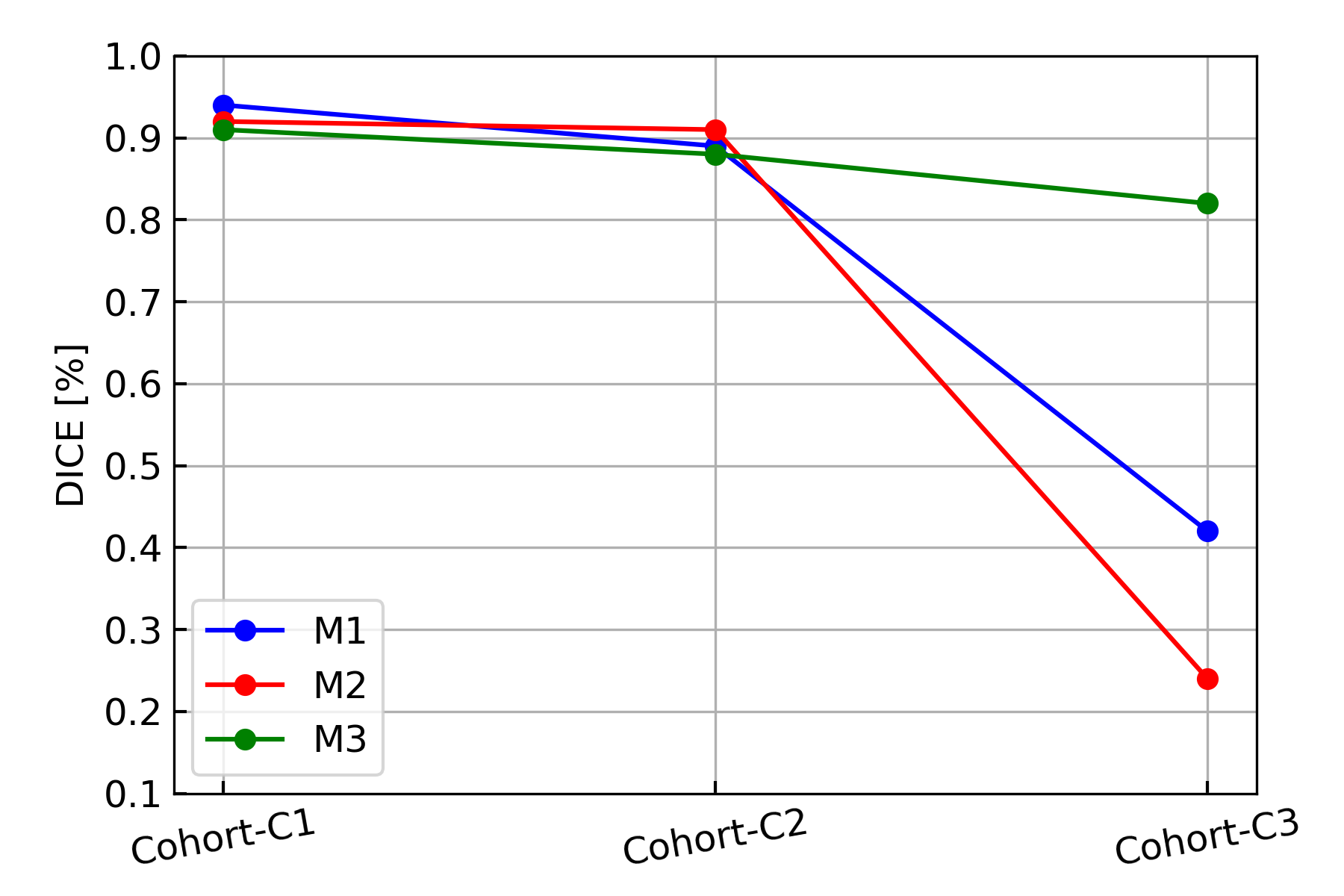}
    \caption{Continual segmentation results for domain generalization. $\mathbf{M}_{s}$: model trained using the train data from cohort C1, $\mathbf{M}_{t_{1}}$: model $\mathbf{M}_{s}$, finetuned using 10 TRUS images from cohort C2, and $\mathbf{M}_{t_{2}}$: model $\mathbf{M}_{t_{1}}$, finetuned using 10 TRUS images from cohort C3.}
    \label{fig:continual}
\end{figure}

When training $\mathbf{M}_{t_{1}}$ by finetuning $\mathbf{M}_{s} $with knowledge distillation loss in the 10 cases from cohort C2, we tested the model $\mathbf{M}_{t_{1}}$ on the studies in cohort C2-test and C3-test, but as well in studies in cohort C1-test to assess whether the model suffered from catastrophic forgetting. Based on the results, the final model ($\mathbf{M}_{t_{2}}$), even though had a small drop of performance on the test studies in cohorts C1 and C2, achieved the overall best performance across all three data cohorts with a Dice score of 0.91, 0.88 and 0.82 for the test studies in cohorts C1, C2 and C3 (Fig. \ref{fig:continual}). 

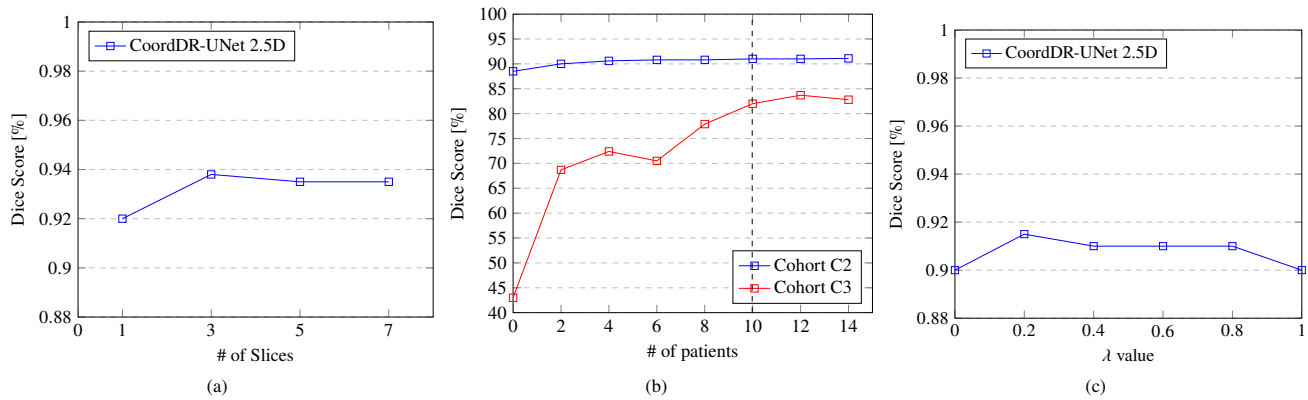
\begin{figure*}[!t]
    \begin{subfigure}[b]{0.31\textwidth}
      \centering
      \resizebox{\linewidth}{!}{
        \begin{tikzpicture}
        \begin{axis}[
            xlabel={\# of Slices},
            ylabel={Dice Score [\%]},
            xmin=0, xmax=8,
            ymin=0.88, ymax=1.0,
            xtick={0,1,3,5,7},
            ytick={0.88, 0.90, 0.92, 0.94, 0.96,0.98, 1.00},
            legend pos=north west,
            ymajorgrids=true,
            grid style=dashed]
            \addplot[color=blue,mark=square,]
            coordinates {(1,0.92)(3,0.938)(5, 0.935)(7, 0.935)};
            \legend{CoordDR-UNet 2.5D}
        \end{axis}
        \end{tikzpicture}}  
        \vspace*{-5mm}
      \caption{}
        \label{fig:ablationA}
    \end{subfigure}
     \begin{subfigure}[b]{0.31\textwidth}
      \centering
      \resizebox{\linewidth}{!}{
         \begin{tikzpicture}
                \begin{axis}[
                    xlabel={\# of patients},
                    ylabel={Dice Score [\%]},
                    xmin=0, xmax=15,
                    ymin=40, ymax=100,
                    xtick={0, 2,4,6,8,10, 12, 14},
                    ytick={40, 45, 50, 55, 60, 65, 70, 75, 80, 85, 90, 95, 100},
                    legend pos=south east,
                    ymajorgrids=true,
                    grid style=dashed,
                ]
                \addplot[
                    color=blue,
                    mark=square,
                    ]
                    coordinates {
                    (0,88.5)(2,90.0)(4,90.6)(6, 90.8)(8,90.8)(10,91.0)(12, 91.0)(14, 91.1)
                    };
                \addplot[
                    color=red,
                    mark=square,
                    ]
                    coordinates {
                    (0,43.0)(2,68.7)(4,72.4)(6, 70.5)(8, 77.9)(10, 82.0)(12, 83.7)(14, 82.8)
                    };
                \addlegendentry{Cohort C2}
                \addlegendentry{Cohort C3}
                \end{axis}
                \draw [dashed] (4.56,0) -- (4.56,5.6);
            \end{tikzpicture}} 
          \vspace*{-5mm}
        \caption{}
     \label{fig:ablationB}
     \end{subfigure}
     \begin{subfigure}[b]{0.31\textwidth}
        \centering
         \resizebox{\linewidth}{!}{
        \begin{tikzpicture}
            \begin{axis}[
                xlabel={$\lambda$ value},
                ylabel={Dice Score [\%]},
                xmin=0.0, xmax=1.0,
                ymin=0.88, ymax=1.0,
                xtick={0.0, 0.2, 0.4, 0.6, 0.8, 1.0},
                ytick={0.88, 0.90, 0.92, 0.94, 0.96,0.98, 1.0},
                legend pos=north west,
                ymajorgrids=true,
                grid style=dashed,
            ]
            \addplot[color=blue, mark=square,]
                coordinates {(0.0,0.90)(0.2,0.915)(0.4, 0.910)(0.6, 0.910)(0.8, 0.91)(1.0, 0.90)};
                \legend{CoordDR-UNet 2.5D}
            \end{axis}
        \end{tikzpicture}}  
      \vspace*{-5mm}

     \caption{}
     \label{fig:ablationC}
     \end{subfigure}
     \label{fig:ablation}
     \caption{(a) CoordDR-UNet 2.5 performance with different number of slices on cohort C1-test set. (b) CoordDR-UNet + LKD finetuning performance with different number of cases on cohort C2-test and C3-test data. (c) CoordDR-UNet + LKD performance with different $\lambda$ values on cohort C2-test data. }
\end{figure*}

    \subsection{Ablation Study}
    \textbf{Impact of Input Data:} We conducted an ablation study and trained the CoordDR-UNet 2.5D with a different number of slices $c \in [1, 3, 5, 7]$ as the input to determine the optimal number of neighboring slices. Fig. \ref{fig:ablationA} shows the Dice score on cohort C1-test data for the different values of $c$. The model trained with a single (2D) slice as input and no neighboring slices obtained a Dice score of 0.92. However, with three slices as input data, the CoordDR-UNet 2.5 achieved a Dice score of 0.94. Furthermore, increasing the number of neighboring slices in the CoordDR-UNet 2.5 model did not improve the segmentation accuracy and slightly decreased the Dice score. Therefore, we trained all 2.5D models with three slices as input data. Also, previous work on prostate gland segmentation in MRI \citep{simon2021} showed that having a 2.5D model with three adjacent slices is successful in generating accurate segmentation on prostate MRI.\\
    \noindent  \textbf{Impact of Finetuning Cases:}  To evaluate the impact of training data size for model finetuning and its influence on domain generalization across cohort C2 and C3 test data, we conducted experiments by finetuning CoordDR-UNet+KDL with the various number of cases $I \in [0, 2, 4, 6, 8, 10, 12, 14]$. Fig. \ref{fig:ablationB} shows the segmentation Dice coefficient score of CoordDR-UNet+KDL for cohort C2-test and C3-test. The model finetuned with two subjects achieved a Dice score of 0.90, and adding more cases, only slightly improved the segmentation performance (increase to 0.91). However, finetuning with 10, 12, and 14 subjects have a similar Dice score of 0.91. These results imply that when TRUS images from a new institution are acquired using a similar ultrasound vendor and probe to training data, just a few instances are required to achieve satisfactory prostate gland segmentation.  For cohort C3-test data, finetuning with more data showed a significant improvement in Dice score. Since the TRUS images in this cohort are significantly different and acquired using a side-fire probe. The model without any finetuning subject achieved a Dice score of 0.43, however, with adding more cases, the Dice score increased. For instance, the model finetuned with 12 subjects achieved a Dice score of 0.84.\\
    \noindent \textbf{Impact of Lambda:} To control the influence of knowledge distillation loss during finetuning, we used $\lambda$ as a weight factor. If we set $\lambda =0$, then the model is trained using the classical finetuning scenario with no knowledge distillation. To find out the impact of $\lambda$, we finetuned CoordDR-UNet+KDL model on cohort C2 training data (10 cases) with various $\lambda$ values. Fig. \ref{fig:ablationC} shows the line-plot and the computed Dice value for a range of $\lambda$ values $[0 - 1.0]$. The value of $\lambda =0.2$ achieved the best Dice value on the test data and increasing the $\lambda$ resulted in a reduction in segmentation accuracy. Therefore, we set the $\lambda =0.2$ for the rest of the experiments.

\section{Discussion}
In this study, we introduced a deep learning model for prostate gland segmentation in 3D TRUS scans called Coordination Dilated-Residual UNet (CoordDR-UNet), as well as a strategy for domain generalization which was tested using TRUS images from three different institutions. Our experiments demonstrated that the proposed method accurately localized and segmented the prostate gland in the presence of variations in image acquisition parameters and generalized well on data from multiple institutions. These encouraging results are attributable to the addition of a coordination attention block in our model that increased segmentation accuracy in ambiguous regions (e.g., prostate borders at the apex). Moreover, we introduced a model generalization approach for prostate gland segmentation in 3D TRUS based on distilling knowledge from previously trained models. The primary goal of using the knowledge distillation strategy during finetuning was to reduce the impact of catastrophic forgetting, which manifests as a performance loss after each finetuning cycle on data from a new center. The experiment results showed (Table 3) that when the CoordDR-UNet was finetuned on cohort C3 using the classical finetuning approach, it obtained a Dice score of $0.80\pm0.03$ (P$\leq$0.05) and HD95 value of 7.30 $mm$, whereas the CoordDR-UNet+KDL approach increased the Dice score by 2.0\% ($0.82\pm 0.16$. p$\leq$0.05). Similarly, for cohort C2, the CoordDR-UNet+KDL outperformed the classical finetuning approach.

Furthermore, we compared our method with existing segmentation methods. While a direct comparison with published results from prior studies was not feasible due to a lack of data availability, we trained the prior algorithms on our cohort and designed fair comparative experiments. Our proposed model outperformed other models in several evaluation metrics (e.g., the boundary distance between the prediction and ground-truth for all cases in the test cohorts was on average only 4.37 $mm$ versus 7.9 $mm$ for alternative approaches). 

Our proposed generalization framework has both similarities and differences to federated learning approaches \citep{MAL-083, Liu_2021_CVPR}. Similar to federated learning,  we transferred our learned model weights and training code to an institution without directly accessing their data. The model was then trained, finetuned, and tested locally, with the results being reported back. Yet, unlike federated learning, we did not apply privacy-preserving techniques, and the weights aggregation was solely based on retraining with new data. Further work will involve testing the model on data from additional institutions without data sharing. 

Our study has four limitations. First, we only segmented the whole prostate gland without segmenting the distinct anatomic zones, which have the potential for other tasks beyond TRUS-MRI fusion. Second, our study includes images from only two ultrasound vendors. Future studies will expand this work to include TRUS scans from other ultrasound vendors. Third, while our proposed method achieved encouraging results on the studies in cohort C3-test, there is still potential for further improvement. Adopting additional data augmentation strategies might assist in addressing issues related to variations in field of view and ultrasound probes such as side-fire and end-fire. Fourth, the number of samples chosen from C2 and C3 (10 patients) to finetune the models can be ablated to find the optimal number of patients for better segmentation accuracy.

Our proposed approach for automated prostate segmentation on TRUS images can improve clinical workflow in four ways. First, our approach aids in the diagnosis of prostate cancer by enabling a better biopsy procedure with more accurately registered MRI-TRUS images and therefore possibly better needle targeting during biopsy (the registrations are driven by the segmentation of the prostate provided by our approach). The MRI-TRUS fusion step during the targeted biopsy is prone to errors \citep{Das2020, Avola2021}, and registration accuracy is affected by the quality of prostate gland segmentation on both TRUS and MR images. Second, our generalizable approach is capable of producing high-quality segmentation for a wide range of probes, and it can be integrated into TRUS scanners as an alternative method for prostate segmentation. Third, our approach provides objective, reproducible, and fast estimates of prostate volume (run-time including pre- and post-processing steps: 12 seconds), which has been reported to be operator-dependent, difficult to replicate, and less accurate \citep{vanSloun2021}. Fourth, our approach allows for better planning of focal treatment to mark the prostate capsule while sparing the tissue beyond the prostate.

\section{Conclusion}
We have introduced an accurate and generalizable approach for prostate segmentation in 3D transrectal ultrasound images to limit the need for manual segmentation of the prostate during targeted biopsy procedures. The proposed deep learning framework outperformed state-of-the-art segmentation approaches in accuracy and generalization and was tested on data from three institutions. In comparison to traditional approaches that require substantial user input and processing time, our pipeline delivered accurate and efficient segmentation of the prostate without any user input. The ease of use and speed of our pipeline make it appealing for practical deployment to allow direct segmentation of the prostate during biopsy or treatment procedures. In the future, we would like to evaluate our method on other tasks and imaging modalities, e.g., prostate segmentation in MRI or CT. 

\section{Credit authorship contribution statement}
\textbf{Sulaiman Vesal}: Conceptualization, Methodology, Software, Formal analysis, Investigation, Data curation, Writing an original draft, Visualization.
\textbf{Iani Gayo}: Data curation, Software, Formal analysis \& Visualization. 
\textbf{Indrani Bhattacharya}: Formal analysis, Writing review \& editing. 
\textbf{Shyam Natarajan}: Data curation, Resources, Writing review \& editing. 
\textbf{Leonard S. Marks}: Data curation, Writing review \& editing. 
\textbf{Dean C Barratt}: Data curation, Writing review \& editing. 
\textbf{Richard E. Fan}: Resources, Writing review \& editing.
\textbf{Yipeng Hu}: Data curation, Resources, Writing review \& editing. 
\textbf{Geoffrey A. Sonn}: Resources, Project administration, Funding acquisition, Supervision, Data curation, Writing review \& editing.
\textbf{Mirabela Rusu}: Conceptualization, Project administration, Methodology, Resources,  Writing review \& editing, Supervision.

\section*{Declaration of Competing Interest}
The authors declare that they do not have any competing financial interests or personal relationships that could have influenced the work in this paper.

\section*{Acknowledgments}
We acknowledge the following funding sources: Departments of Radiology and Urology, Stanford University and  International Alliance for Cancer Early Detection (ACED).
Research reported in this publication was supported by the National Cancer Institute of the National Institutes of Health under Award Number R37CA260346. The content is solely the responsibility of the authors and does not necessarily represent the official views of the National Institutes of Health.

\bibliographystyle{model2-names.bst}\biboptions{authoryear}
\bibliography{refs}

\begin{thebibliography}{62}
\expandafter\ifx\csname natexlab\endcsname\relax\def\natexlab#1{#1}\fi
\providecommand{\url}[1]{\texttt{#1}}
\providecommand{\href}[2]{#2}
\providecommand{\path}[1]{#1}
\providecommand{\DOIprefix}{doi:}
\providecommand{\ArXivprefix}{arXiv:}
\providecommand{\URLprefix}{URL: }
\providecommand{\Pubmedprefix}{pmid:}
\providecommand{\doi}[1]{\href{http://dx.doi.org/#1}{\path{#1}}}
\providecommand{\Pubmed}[1]{\href{pmid:#1}{\path{#1}}}
\providecommand{\bibinfo}[2]{#2}
\ifx\xfnm\relax \def\xfnm[#1]{\unskip,\space#1}\fi
\bibitem[{Aldoj et~al.(2020)Aldoj, Biavati, Michallek, Stober and
  Dewey}]{Aldoj2020}
\bibinfo{author}{Aldoj, N.}, \bibinfo{author}{Biavati, F.},
  \bibinfo{author}{Michallek, F.}, \bibinfo{author}{Stober, S.},
  \bibinfo{author}{Dewey, M.}, \bibinfo{year}{2020}.
\newblock \bibinfo{title}{Automatic prostate and prostate zones segmentation of
  magnetic resonance images using densenet-like u-net}.
\newblock \bibinfo{journal}{Scientific Reports} \bibinfo{volume}{10},
  \bibinfo{pages}{14315}.
\bibitem[{Anas et~al.(2018)Anas, Mousavi and Abolmaesumi}]{ANAS2018107}
\bibinfo{author}{Anas, E.M.A.}, \bibinfo{author}{Mousavi, P.},
  \bibinfo{author}{Abolmaesumi, P.}, \bibinfo{year}{2018}.
\newblock \bibinfo{title}{A deep learning approach for real time prostate
  segmentation in freehand ultrasound guided biopsy}.
\newblock \bibinfo{journal}{Medical Image Analysis} \bibinfo{volume}{48},
  \bibinfo{pages}{107--116}.
\bibitem[{Anas et~al.(2017)Anas, Nouranian, Mahdavi, Spadinger, Morris,
  Salcudean, Mousavi and Abolmaesumi}]{Ans2017}
\bibinfo{author}{Anas, E.M.A.}, \bibinfo{author}{Nouranian, S.},
  \bibinfo{author}{Mahdavi, S.S.}, \bibinfo{author}{Spadinger, I.},
  \bibinfo{author}{Morris, W.J.}, \bibinfo{author}{Salcudean, S.E.},
  \bibinfo{author}{Mousavi, P.}, \bibinfo{author}{Abolmaesumi, P.},
  \bibinfo{year}{2017}.
\newblock \bibinfo{title}{Clinical target-volume delineation in prostate
  brachytherapy using residual neural networks}, in:
  \bibinfo{booktitle}{Medical Image Computing and Computer Assisted
  Intervention, MICCAI 2017}, pp. \bibinfo{pages}{365--373}.
\bibitem[{Avola et~al.(2021)Avola, Cinque, Fagioli, Foresti and
  Mecca}]{Avola2021}
\bibinfo{author}{Avola, D.}, \bibinfo{author}{Cinque, L.},
  \bibinfo{author}{Fagioli, A.}, \bibinfo{author}{Foresti, G.},
  \bibinfo{author}{Mecca, A.}, \bibinfo{year}{2021}.
\newblock \bibinfo{title}{Ultrasound medical imaging techniques: A survey}.
\newblock \bibinfo{journal}{ACM Comput. Surv.} \bibinfo{volume}{54}.
\bibitem[{Azizi et~al.(2018)Azizi, Van~Woudenberg, Sojoudi, Li, Xu, Abu~Anas,
  Yan, Tahmasebi, Kwak, Turkbey, Choyke, Pinto, Wood, Mousavi and
  Abolmaesumi}]{Azizi2018}
\bibinfo{author}{Azizi, S.}, \bibinfo{author}{Van~Woudenberg, N.},
  \bibinfo{author}{Sojoudi, S.}, \bibinfo{author}{Li, M.}, \bibinfo{author}{Xu,
  S.}, \bibinfo{author}{Abu~Anas, E.M.}, \bibinfo{author}{Yan, P.},
  \bibinfo{author}{Tahmasebi, A.}, \bibinfo{author}{Kwak, J.T.},
  \bibinfo{author}{Turkbey, B.}, \bibinfo{author}{Choyke, P.},
  \bibinfo{author}{Pinto, P.}, \bibinfo{author}{Wood, B.},
  \bibinfo{author}{Mousavi, P.}, \bibinfo{author}{Abolmaesumi, P.},
  \bibinfo{year}{2018}.
\newblock \bibinfo{title}{Toward a real-time system for temporal enhanced
  ultrasound-guided prostate biopsy}.
\newblock \bibinfo{journal}{International Journal of Computer Assisted
  Radiology and Surgery} \bibinfo{volume}{13}, \bibinfo{pages}{1201--1209}.
\bibitem[{Cermelli et~al.(2020)Cermelli, Mancini, Rota~Bulò, Ricci and
  Caputo}]{CL_Mib}
\bibinfo{author}{Cermelli, F.}, \bibinfo{author}{Mancini, M.},
  \bibinfo{author}{Rota~Bulò, S.}, \bibinfo{author}{Ricci, E.},
  \bibinfo{author}{Caputo, B.}, \bibinfo{year}{2020}.
\newblock \bibinfo{title}{Modeling the background for incremental learning in
  semantic segmentation}, in: \bibinfo{booktitle}{2020 IEEE/CVF Conference on
  Computer Vision and Pattern Recognition (CVPR)}, pp.
  \bibinfo{pages}{9230--9239}.
\bibitem[{Clark et~al.(2013)Clark, Vendt, Smith, Freymann, Kirby, Koppel,
  Moore, Phillips, Maffitt, Pringle, Tarbox and Prior}]{Clark2013}
\bibinfo{author}{Clark, K.}, \bibinfo{author}{Vendt, B.},
  \bibinfo{author}{Smith, K.}, \bibinfo{author}{Freymann, J.},
  \bibinfo{author}{Kirby, J.}, \bibinfo{author}{Koppel, P.},
  \bibinfo{author}{Moore, S.}, \bibinfo{author}{Phillips, S.},
  \bibinfo{author}{Maffitt, D.}, \bibinfo{author}{Pringle, M.},
  \bibinfo{author}{Tarbox, L.}, \bibinfo{author}{Prior, F.},
  \bibinfo{year}{2013}.
\newblock \bibinfo{title}{The cancer imaging archive (tcia): Maintaining and
  operating a public information repository}.
\newblock \bibinfo{journal}{Journal of Digital Imaging} \bibinfo{volume}{26},
  \bibinfo{pages}{1045--1057}.
\bibitem[{Das et~al.(2020)Das, Razik, Netaji and Verma}]{Das2020}
\bibinfo{author}{Das, C.J.}, \bibinfo{author}{Razik, A.},
  \bibinfo{author}{Netaji, A.}, \bibinfo{author}{Verma, S.},
  \bibinfo{year}{2020}.
\newblock \bibinfo{title}{Prostate mri--trus fusion biopsy: a review of the
  state of the art procedure}.
\newblock \bibinfo{journal}{Abdominal Radiology} \bibinfo{volume}{45},
  \bibinfo{pages}{2176--2183}.
\bibitem[{Feng et~al.(2021)Feng, You, Chen, Zhang, Zhu, Wu, Wu and
  Chen}]{pmlr-v139-feng21f}
\bibinfo{author}{Feng, H.}, \bibinfo{author}{You, Z.}, \bibinfo{author}{Chen,
  M.}, \bibinfo{author}{Zhang, T.}, \bibinfo{author}{Zhu, M.},
  \bibinfo{author}{Wu, F.}, \bibinfo{author}{Wu, C.}, \bibinfo{author}{Chen,
  W.}, \bibinfo{year}{2021}.
\newblock \bibinfo{title}{Kd3a: Unsupervised multi-source decentralized domain
  adaptation via knowledge distillation}, in: \bibinfo{editor}{Meila, M.},
  \bibinfo{editor}{Zhang, T.} (Eds.), \bibinfo{booktitle}{Proceedings of the
  38th International Conference on Machine Learning},
  \bibinfo{publisher}{PMLR}. pp. \bibinfo{pages}{3274--3283}.
\bibitem[{Ghavami et~al.(2018)Ghavami, Hu, Bonmati, Rodell, Gibson, Moore and
  Barratt}]{Ghavami2018}
\bibinfo{author}{Ghavami, N.}, \bibinfo{author}{Hu, Y.},
  \bibinfo{author}{Bonmati, E.}, \bibinfo{author}{Rodell, R.},
  \bibinfo{author}{Gibson, E.}, \bibinfo{author}{Moore, C.},
  \bibinfo{author}{Barratt, D.}, \bibinfo{year}{2018}.
\newblock \bibinfo{title}{{Integration of spatial information in convolutional
  neural networks for automatic segmentation of intraoperative transrectal
  ultrasound images}}.
\newblock \bibinfo{journal}{Journal of Medical Imaging} \bibinfo{volume}{6},
  \bibinfo{pages}{1 -- 6}.
\bibitem[{Ghose et~al.(2012)Ghose, Oliver, Martí, Lladó, Vilanova, Freixenet,
  Mitra, Sidibé and Meriaudeau}]{GHOSE2012262}
\bibinfo{author}{Ghose, S.}, \bibinfo{author}{Oliver, A.},
  \bibinfo{author}{Martí, R.}, \bibinfo{author}{Lladó, X.},
  \bibinfo{author}{Vilanova, J.C.}, \bibinfo{author}{Freixenet, J.},
  \bibinfo{author}{Mitra, J.}, \bibinfo{author}{Sidibé, D.},
  \bibinfo{author}{Meriaudeau, F.}, \bibinfo{year}{2012}.
\newblock \bibinfo{title}{A survey of prostate segmentation methodologies in
  ultrasound, magnetic resonance and computed tomography images}.
\newblock \bibinfo{journal}{Computer Methods and Programs in Biomedicine}
  \bibinfo{volume}{108}, \bibinfo{pages}{262--287}.
\bibitem[{Girum et~al.(2020)Girum, Lalande, Hussain and
  Cr{\'e}hange}]{Girum2020}
\bibinfo{author}{Girum, K.B.}, \bibinfo{author}{Lalande, A.},
  \bibinfo{author}{Hussain, R.}, \bibinfo{author}{Cr{\'e}hange, G.},
  \bibinfo{year}{2020}.
\newblock \bibinfo{title}{A deep learning method for real-time intraoperative
  us image segmentation in prostate brachytherapy}.
\newblock \bibinfo{journal}{International Journal of Computer Assisted
  Radiology and Surgery} \bibinfo{volume}{15}, \bibinfo{pages}{1467--1476}.
\bibitem[{Harvey et~al.(2012)Harvey, Pilcher, Richenberg, Patel and
  Frauscher}]{Harvey2012}
\bibinfo{author}{Harvey, C.J.}, \bibinfo{author}{Pilcher, J.},
  \bibinfo{author}{Richenberg, J.}, \bibinfo{author}{Patel, U.},
  \bibinfo{author}{Frauscher, F.}, \bibinfo{year}{2012}.
\newblock \bibinfo{title}{Applications of transrectal ultrasound in prostate
  cancer}.
\newblock \bibinfo{journal}{The British Journal of Radiology}
  \bibinfo{volume}{85}, \bibinfo{pages}{S3--S17}.
\bibitem[{He et~al.(2016)He, Zhang, Ren and Sun}]{resdeep}
\bibinfo{author}{He, K.}, \bibinfo{author}{Zhang, X.}, \bibinfo{author}{Ren,
  S.}, \bibinfo{author}{Sun, J.}, \bibinfo{year}{2016}.
\newblock \bibinfo{title}{Deep residual learning for image recognition}, in:
  \bibinfo{booktitle}{2016 IEEE Conference on Computer Vision and Pattern
  Recognition (CVPR)}, pp. \bibinfo{pages}{770--778}.
\bibitem[{Hinton et~al.(2015)Hinton, Vinyals and Dean}]{hinton2015}
\bibinfo{author}{Hinton, G.}, \bibinfo{author}{Vinyals, O.},
  \bibinfo{author}{Dean, J.}, \bibinfo{year}{2015}.
\newblock \bibinfo{title}{Distilling the knowledge in a neural network}, in:
  \bibinfo{booktitle}{NIPS Deep Learning and Representation Learning Workshop}.
\bibitem[{Hou et~al.(2021)Hou, Zhou and Feng}]{Hou_2021_CVPR}
\bibinfo{author}{Hou, Q.}, \bibinfo{author}{Zhou, D.}, \bibinfo{author}{Feng,
  J.}, \bibinfo{year}{2021}.
\newblock \bibinfo{title}{Coordinate attention for efficient mobile network
  design}, in: \bibinfo{booktitle}{Proceedings of the IEEE/CVF Conference on
  Computer Vision and Pattern Recognition (CVPR)}, pp.
  \bibinfo{pages}{13713--13722}.
\bibitem[{Hu et~al.(2018)Hu, Shen and Sun}]{8578843}
\bibinfo{author}{Hu, J.}, \bibinfo{author}{Shen, L.}, \bibinfo{author}{Sun,
  G.}, \bibinfo{year}{2018}.
\newblock \bibinfo{title}{Squeeze-and-excitation networks}, in:
  \bibinfo{booktitle}{2018 IEEE/CVF Conference on Computer Vision and Pattern
  Recognition}, pp. \bibinfo{pages}{7132--7141}.
\bibitem[{Jaouen et~al.(2019)Jaouen, Bert, Mountris, Boussion, Schick, Pradier,
  Valeri and Visvikis}]{Jaouen2019}
\bibinfo{author}{Jaouen, V.}, \bibinfo{author}{Bert, J.},
  \bibinfo{author}{Mountris, K.A.}, \bibinfo{author}{Boussion, N.},
  \bibinfo{author}{Schick, U.}, \bibinfo{author}{Pradier, O.},
  \bibinfo{author}{Valeri, A.}, \bibinfo{author}{Visvikis, D.},
  \bibinfo{year}{2019}.
\newblock \bibinfo{title}{Prostate volume segmentation in trus using hybrid
  edge-bhattacharyya active surfaces}.
\newblock \bibinfo{journal}{IEEE Transactions on Biomedical Engineering}
  \bibinfo{volume}{66}, \bibinfo{pages}{920--933}.
\bibitem[{Jung et~al.(2020)Jung, Wada, Crall, Tanaka, Graving, Reinders, Yadav,
  Banerjee, Vecsei, Kraft, Rui, Borovec, Vallentin, Zhydenko, Pfeiffer, Cook,
  Fernández, De~Rainville, Weng, Ayala-Acevedo, Meudec, Laporte
  et~al.}]{imgaug}
\bibinfo{author}{Jung, A.B.}, \bibinfo{author}{Wada, K.},
  \bibinfo{author}{Crall, J.}, \bibinfo{author}{Tanaka, S.},
  \bibinfo{author}{Graving, J.}, \bibinfo{author}{Reinders, C.},
  \bibinfo{author}{Yadav, S.}, \bibinfo{author}{Banerjee, J.},
  \bibinfo{author}{Vecsei, G.}, \bibinfo{author}{Kraft, A.},
  \bibinfo{author}{Rui, Z.}, \bibinfo{author}{Borovec, J.},
  \bibinfo{author}{Vallentin, C.}, \bibinfo{author}{Zhydenko, S.},
  \bibinfo{author}{Pfeiffer, K.}, \bibinfo{author}{Cook, B.},
  \bibinfo{author}{Fernández, I.}, \bibinfo{author}{De~Rainville, F.M.},
  \bibinfo{author}{Weng, C.H.}, \bibinfo{author}{Ayala-Acevedo, A.},
  \bibinfo{author}{Meudec, R.}, \bibinfo{author}{Laporte, M.}, et~al.,
  \bibinfo{year}{2020}.
\newblock \bibinfo{title}{{imgaug}}.
\newblock \bibinfo{howpublished}{\url{https://github.com/aleju/imgaug}}.
\newblock \bibinfo{note}{Online; accessed 01-Sep-2021}.
\bibitem[{Kairouz et~al.(2021)Kairouz, McMahan, Avent, Bellet, Bennis and
  et~al.}]{MAL-083}
\bibinfo{author}{Kairouz, P.}, \bibinfo{author}{McMahan, H.B.},
  \bibinfo{author}{Avent, B.}, \bibinfo{author}{Bellet, A.},
  \bibinfo{author}{Bennis, M.}, \bibinfo{author}{et~al., A.N.B.},
  \bibinfo{year}{2021}.
\newblock \bibinfo{title}{Advances and open problems in federated learning}.
\newblock \bibinfo{journal}{Foundations and Trends® in Machine Learning}
  \bibinfo{volume}{14}, \bibinfo{pages}{1--210}.
\bibitem[{Karimi et~al.(2019)Karimi, Zeng, Mathur, Avinash, Mahdavi, Spadinger,
  Abolmaesumi and Salcudean}]{KARIMI2019186}
\bibinfo{author}{Karimi, D.}, \bibinfo{author}{Zeng, Q.},
  \bibinfo{author}{Mathur, P.}, \bibinfo{author}{Avinash, A.},
  \bibinfo{author}{Mahdavi, S.}, \bibinfo{author}{Spadinger, I.},
  \bibinfo{author}{Abolmaesumi, P.}, \bibinfo{author}{Salcudean, S.E.},
  \bibinfo{year}{2019}.
\newblock \bibinfo{title}{Accurate and robust deep learning-based segmentation
  of the prostate clinical target volume in ultrasound images}.
\newblock \bibinfo{journal}{Medical Image Analysis} \bibinfo{volume}{57},
  \bibinfo{pages}{186--196}.
\bibitem[{Kim et~al.(2021)Kim, Yoo, Park, Kim and Lee}]{kim2021selfreg}
\bibinfo{author}{Kim, D.}, \bibinfo{author}{Yoo, Y.}, \bibinfo{author}{Park,
  S.}, \bibinfo{author}{Kim, J.}, \bibinfo{author}{Lee, J.},
  \bibinfo{year}{2021}.
\newblock \bibinfo{title}{Selfreg: Self-supervised contrastive regularization
  for domain generalization}, in: \bibinfo{booktitle}{Proceedings of the
  IEEE/CVF International Conference on Computer Vision}, pp.
  \bibinfo{pages}{9619--9628}.
\bibitem[{Lei et~al.(2019)Lei, Tian, He, Wang, Wang, Patel, Jani, Mao, Curran,
  Liu and Yang}]{Lei2019}
\bibinfo{author}{Lei, Y.}, \bibinfo{author}{Tian, S.}, \bibinfo{author}{He,
  X.}, \bibinfo{author}{Wang, T.}, \bibinfo{author}{Wang, B.},
  \bibinfo{author}{Patel, P.}, \bibinfo{author}{Jani, A.B.},
  \bibinfo{author}{Mao, H.}, \bibinfo{author}{Curran, W.J.},
  \bibinfo{author}{Liu, T.}, \bibinfo{author}{Yang, X.}, \bibinfo{year}{2019}.
\newblock \bibinfo{title}{Ultrasound prostate segmentation based on
  multidirectional deeply supervised v-net}.
\newblock \bibinfo{journal}{Medical Physics} \bibinfo{volume}{46},
  \bibinfo{pages}{3194--3206}.
\bibitem[{Li et~al.(2016)Li, Li, Fedorov, Kapur and Yang}]{li2016}
\bibinfo{author}{Li, X.}, \bibinfo{author}{Li, C.}, \bibinfo{author}{Fedorov,
  A.}, \bibinfo{author}{Kapur, T.}, \bibinfo{author}{Yang, X.},
  \bibinfo{year}{2016}.
\newblock \bibinfo{title}{Segmentation of prostate from ultrasound images using
  level sets on active band and intensity variation across edges}.
\newblock \bibinfo{journal}{Medical Physics} \bibinfo{volume}{43},
  \bibinfo{pages}{3090--3103}.
\bibitem[{Liang et~al.(2022)Liang, Wu, Li, Qin, Zhang and Liu}]{Xiaobo2022}
\bibinfo{author}{Liang, X.}, \bibinfo{author}{Wu, L.}, \bibinfo{author}{Li,
  J.}, \bibinfo{author}{Qin, T.}, \bibinfo{author}{Zhang, M.},
  \bibinfo{author}{Liu, T.Y.}, \bibinfo{year}{2022}.
\newblock \bibinfo{title}{Multi-teacher distillation with single model for
  neural machine translation}.
\newblock \bibinfo{journal}{IEEE/ACM Transactions on Audio, Speech, and
  Language Processing} \bibinfo{volume}{30}, \bibinfo{pages}{992--1002}.
\newblock \DOIprefix\doi{10.1109/TASLP.2022.3153264}.
\bibitem[{Liau et~al.(2019)Liau, Goldberg and Arif-Tiwari}]{Liau2019}
\bibinfo{author}{Liau, J.}, \bibinfo{author}{Goldberg, D.},
  \bibinfo{author}{Arif-Tiwari, H.}, \bibinfo{year}{2019}.
\newblock \bibinfo{title}{Prostate cancer detection and diagnosis: Role of
  ultrasound with mri correlates}.
\newblock \bibinfo{journal}{Current Radiology Reports} \bibinfo{volume}{7},
  \bibinfo{pages}{7}.
\bibitem[{Liu et~al.(2021)Liu, Chen, Qin, Dou and Heng}]{Liu_2021_CVPR}
\bibinfo{author}{Liu, Q.}, \bibinfo{author}{Chen, C.}, \bibinfo{author}{Qin,
  J.}, \bibinfo{author}{Dou, Q.}, \bibinfo{author}{Heng, P.A.},
  \bibinfo{year}{2021}.
\newblock \bibinfo{title}{Feddg: Federated domain generalization on medical
  image segmentation via episodic learning in continuous frequency space}, in:
  \bibinfo{booktitle}{Proceedings of the IEEE/CVF Conference on Computer Vision
  and Pattern Recognition (CVPR)}, pp. \bibinfo{pages}{1013--1023}.
\bibitem[{Liu et~al.(2020)Liu, Dou, Yu and Heng}]{Liu2020}
\bibinfo{author}{Liu, Q.}, \bibinfo{author}{Dou, Q.}, \bibinfo{author}{Yu, L.},
  \bibinfo{author}{Heng, P.A.}, \bibinfo{year}{2020}.
\newblock \bibinfo{title}{Ms-net: Multi-site network for improving prostate
  segmentation with heterogeneous mri data}.
\newblock \bibinfo{journal}{IEEE Transactions on Medical Imaging}
  \bibinfo{volume}{39}, \bibinfo{pages}{2713--2724}.
\bibitem[{Meng et~al.(2018)Meng, Li, Gong and Juang}]{meng2018}
\bibinfo{author}{Meng, Z.}, \bibinfo{author}{Li, J.}, \bibinfo{author}{Gong,
  Y.}, \bibinfo{author}{Juang, B.H.}, \bibinfo{year}{2018}.
\newblock \bibinfo{title}{Adversarial teacher-student learning for unsupervised
  domain adaptation}, in: \bibinfo{booktitle}{2018 IEEE International
  Conference on Acoustics, Speech and Signal Processing (ICASSP)}, pp.
  \bibinfo{pages}{5949--5953}.
\newblock \DOIprefix\doi{10.1109/ICASSP.2018.8461682}.
\bibitem[{Michalski et~al.(2016)Michalski, Pisansky, Lawton and
  Potters}]{MICHALSKI20161038}
\bibinfo{author}{Michalski, J.M.}, \bibinfo{author}{Pisansky, T.M.},
  \bibinfo{author}{Lawton, C.A.}, \bibinfo{author}{Potters, L.},
  \bibinfo{year}{2016}.
\newblock \bibinfo{title}{Chapter 53 - prostate cancer}, in:
  \bibinfo{booktitle}{Clinical Radiation Oncology (Fourth Edition)}.
  \bibinfo{edition}{fourth edition} ed.. \bibinfo{publisher}{Elsevier},
  \bibinfo{address}{Philadelphia}, pp. \bibinfo{pages}{1038--1095.e18}.
\bibitem[{Michieli and Zanuttigh(2019)}]{lkd2019}
\bibinfo{author}{Michieli, U.}, \bibinfo{author}{Zanuttigh, P.},
  \bibinfo{year}{2019}.
\newblock \bibinfo{title}{Incremental learning techniques for semantic
  segmentation}, in: \bibinfo{booktitle}{2019 IEEE/CVF International Conference
  on Computer Vision Workshop (ICCVW)}, pp. \bibinfo{pages}{3205--3212}.
\bibitem[{Nadeau and Bengio(2003)}]{Nadeau2003}
\bibinfo{author}{Nadeau, C.}, \bibinfo{author}{Bengio, Y.},
  \bibinfo{year}{2003}.
\newblock \bibinfo{title}{Inference for the generalization error}.
\newblock \bibinfo{journal}{Machine Learning} \bibinfo{volume}{52},
  \bibinfo{pages}{239--281}.
\bibitem[{Natarajan et~al.(2020)Natarajan, Priester, Margolis, Huang and
  Marks}]{Natarajan2020}
\bibinfo{author}{Natarajan, S.}, \bibinfo{author}{Priester, A.},
  \bibinfo{author}{Margolis, D.}, \bibinfo{author}{Huang, J.},
  \bibinfo{author}{Marks, L.}, \bibinfo{year}{2020}.
\newblock \bibinfo{title}{{Prostate MRI and Ultrasound With Pathology and
  Coordinates of Tracked Biopsy (Prostate-MRI-US-Biopsy)}}.
\newblock \DOIprefix\doi{10.7937/TCIA.2020.A61IOC1A}.
\bibitem[{Orlando et~al.(2020)Orlando, Gillies, Gyacskov and
  Fenster}]{Nathan2020}
\bibinfo{author}{Orlando, N.}, \bibinfo{author}{Gillies, D.J.},
  \bibinfo{author}{Gyacskov, I.}, \bibinfo{author}{Fenster, A.},
  \bibinfo{year}{2020}.
\newblock \bibinfo{title}{{Deep learning-based automatic prostate segmentation
  in 3D transrectal ultrasound images from multiple acquisition geometries and
  systems}}, in: \bibinfo{booktitle}{Medical Imaging 2020: Image-Guided
  Procedures, Robotic Interventions, and Modeling},
  \bibinfo{organization}{International Society for Optics and Photonics}.
  \bibinfo{publisher}{SPIE}. pp. \bibinfo{pages}{651 -- 656}.
\bibitem[{Park et~al.(2019)Park, Kim, Lu and Cho}]{park2019relational}
\bibinfo{author}{Park, W.}, \bibinfo{author}{Kim, D.}, \bibinfo{author}{Lu,
  Y.}, \bibinfo{author}{Cho, M.}, \bibinfo{year}{2019}.
\newblock \bibinfo{title}{Relational knowledge distillation}, in:
  \bibinfo{booktitle}{Proceedings of the IEEE Conference on Computer Vision and
  Pattern Recognition}, pp. \bibinfo{pages}{3967--3976}.
\bibitem[{Paszke et~al.(2019)Paszke, Gross, Massa, Lerer, Bradbury, Chanan,
  Killeen, Lin, Gimelshein, Antiga, Desmaison, Kopf, Yang, DeVito, Raison,
  Tejani, Chilamkurthy, Steiner, Fang, Bai and Chintala}]{NEURIPS2019_9015}
\bibinfo{author}{Paszke, A.}, \bibinfo{author}{Gross, S.},
  \bibinfo{author}{Massa, F.}, \bibinfo{author}{Lerer, A.},
  \bibinfo{author}{Bradbury, J.}, \bibinfo{author}{Chanan, G.},
  \bibinfo{author}{Killeen, T.}, \bibinfo{author}{Lin, Z.},
  \bibinfo{author}{Gimelshein, N.}, \bibinfo{author}{Antiga, L.},
  \bibinfo{author}{Desmaison, A.}, \bibinfo{author}{Kopf, A.},
  \bibinfo{author}{Yang, E.}, \bibinfo{author}{DeVito, Z.},
  \bibinfo{author}{Raison, M.}, \bibinfo{author}{Tejani, A.},
  \bibinfo{author}{Chilamkurthy, S.}, \bibinfo{author}{Steiner, B.},
  \bibinfo{author}{Fang, L.}, \bibinfo{author}{Bai, J.},
  \bibinfo{author}{Chintala, S.}, \bibinfo{year}{2019}.
\newblock \bibinfo{title}{Pytorch: An imperative style, high-performance deep
  learning library}, in: \bibinfo{booktitle}{Advances in Neural Information
  Processing Systems 32}. \bibinfo{publisher}{Curran Associates, Inc.}, pp.
  \bibinfo{pages}{8024--8035}.
\bibitem[{Pizer et~al.(1990)Pizer, Johnston, Ericksen, Yankaskas and
  Muller}]{Pizer1990}
\bibinfo{author}{Pizer, S.}, \bibinfo{author}{Johnston, R.},
  \bibinfo{author}{Ericksen, J.}, \bibinfo{author}{Yankaskas, B.},
  \bibinfo{author}{Muller, K.}, \bibinfo{year}{1990}.
\newblock \bibinfo{title}{Contrast-limited adaptive histogram equalization:
  speed and effectiveness}, in: \bibinfo{booktitle}{[1990] Proceedings of the
  First Conference on Visualization in Biomedical Computing}, pp.
  \bibinfo{pages}{337--345}.
\newblock \DOIprefix\doi{10.1109/VBC.1990.109340}.
\bibitem[{Ronneberger et~al.(2015)Ronneberger, Fischer and Brox}]{Unet2015}
\bibinfo{author}{Ronneberger, O.}, \bibinfo{author}{Fischer, P.},
  \bibinfo{author}{Brox, T.}, \bibinfo{year}{2015}.
\newblock \bibinfo{title}{U-net: Convolutional networks for biomedical image
  segmentation}, in: \bibinfo{booktitle}{Medical Image Computing and
  Computer-Assisted Intervention -- MICCAI 2015}, pp.
  \bibinfo{pages}{234--241}.
\bibitem[{Roy et~al.(2019)Roy, Navab and Wachinger}]{Roy_TMI2018}
\bibinfo{author}{Roy, A.G.}, \bibinfo{author}{Navab, N.},
  \bibinfo{author}{Wachinger, C.}, \bibinfo{year}{2019}.
\newblock \bibinfo{title}{Recalibrating fully convolutional networks with
  spatial and channel “squeeze and excitation” blocks}.
\newblock \bibinfo{journal}{IEEE Transactions on Medical Imaging}
  \bibinfo{volume}{38}, \bibinfo{pages}{540--549}.
\bibitem[{Sarkar and Das(2016)}]{Sarkar2016}
\bibinfo{author}{Sarkar, S.}, \bibinfo{author}{Das, S.}, \bibinfo{year}{2016}.
\newblock \bibinfo{title}{A review of imaging methods for prostate cancer
  detection: Supplementary issue: Image and video acquisition and processing
  for clinical applications}.
\newblock \bibinfo{journal}{Biomedical Engineering and Computational Biology}
  \bibinfo{volume}{7s1}, \bibinfo{pages}{BECB.S34255}.
\bibitem[{Schimmöller et~al.(2016)Schimmöller, Blondin, Arsov, Rabenalt,
  Albers, Antoch and Quentin}]{Schimmoller2016}
\bibinfo{author}{Schimmöller, L.}, \bibinfo{author}{Blondin, D.},
  \bibinfo{author}{Arsov, C.}, \bibinfo{author}{Rabenalt, R.},
  \bibinfo{author}{Albers, P.}, \bibinfo{author}{Antoch, G.},
  \bibinfo{author}{Quentin, M.}, \bibinfo{year}{2016}.
\newblock \bibinfo{title}{Mri-guided in-bore biopsy: Differences between
  prostate cancer detection and localization in primary and secondary biopsy
  settings}.
\newblock \bibinfo{journal}{American Journal of Roentgenology}
  \bibinfo{volume}{206}, \bibinfo{pages}{92--99}.
\bibitem[{Schlemper et~al.(2019)Schlemper, Oktay, Schaap, Heinrich, Kainz,
  Glocker and Rueckert}]{attunet2019}
\bibinfo{author}{Schlemper, J.}, \bibinfo{author}{Oktay, O.},
  \bibinfo{author}{Schaap, M.}, \bibinfo{author}{Heinrich, M.},
  \bibinfo{author}{Kainz, B.}, \bibinfo{author}{Glocker, B.},
  \bibinfo{author}{Rueckert, D.}, \bibinfo{year}{2019}.
\newblock \bibinfo{title}{Attention gated networks: Learning to leverage
  salient regions in medical images}.
\newblock \bibinfo{journal}{Medical Image Analysis} \bibinfo{volume}{53},
  \bibinfo{pages}{197--207}.
\bibitem[{van Sloun et~al.(2021)van Sloun, Wildeboer, Mannaerts, Postema,
  Gayet, Beerlage, Salomon, Wijkstra and Mischi}]{vanSloun2021}
\bibinfo{author}{van Sloun, R.J.}, \bibinfo{author}{Wildeboer, R.R.},
  \bibinfo{author}{Mannaerts, C.K.}, \bibinfo{author}{Postema, A.W.},
  \bibinfo{author}{Gayet, M.}, \bibinfo{author}{Beerlage, H.P.},
  \bibinfo{author}{Salomon, G.}, \bibinfo{author}{Wijkstra, H.},
  \bibinfo{author}{Mischi, M.}, \bibinfo{year}{2021}.
\newblock \bibinfo{title}{Deep learning for real-time, automatic, and
  scanner-adapted prostate (zone) segmentation of transrectal ultrasound, for
  example, magnetic resonance imaging; transrectal ultrasound fusion prostate
  biopsy}.
\newblock \bibinfo{journal}{European Urology Focus} \bibinfo{volume}{7},
  \bibinfo{pages}{78--85}.
\bibitem[{Soerensen et~al.(2021)Soerensen, Fan, Seetharaman, Chen, Shao,
  Bhattacharya, hun Kim, Sood, Borre, Chung, To'o, Rusu and Sonn}]{simon2021}
\bibinfo{author}{Soerensen, S.J.C.}, \bibinfo{author}{Fan, R.E.},
  \bibinfo{author}{Seetharaman, A.}, \bibinfo{author}{Chen, L.},
  \bibinfo{author}{Shao, W.}, \bibinfo{author}{Bhattacharya, I.},
  \bibinfo{author}{hun Kim, Y.}, \bibinfo{author}{Sood, R.},
  \bibinfo{author}{Borre, M.}, \bibinfo{author}{Chung, B.I.},
  \bibinfo{author}{To'o, K.J.}, \bibinfo{author}{Rusu, M.},
  \bibinfo{author}{Sonn, G.A.}, \bibinfo{year}{2021}.
\newblock \bibinfo{title}{Deep learning improves speed and accuracy of prostate
  gland segmentations on magnetic resonance imaging for targeted biopsy}.
\newblock \bibinfo{journal}{Journal of Urology} \bibinfo{volume}{0},
  \bibinfo{pages}{10.1097/JU.0000000000001783}.
\bibitem[{Sonn et~al.(2013)Sonn, Natarajan, Margolis, MacAiran, Lieu, Huang,
  Dorey and Marks}]{SONN201386}
\bibinfo{author}{Sonn, G.A.}, \bibinfo{author}{Natarajan, S.},
  \bibinfo{author}{Margolis, D.J.}, \bibinfo{author}{MacAiran, M.},
  \bibinfo{author}{Lieu, P.}, \bibinfo{author}{Huang, J.},
  \bibinfo{author}{Dorey, F.J.}, \bibinfo{author}{Marks, L.S.},
  \bibinfo{year}{2013}.
\newblock \bibinfo{title}{Targeted biopsy in the detection of prostate cancer
  using an office based magnetic resonance ultrasound fusion device}.
\newblock \bibinfo{journal}{The Journal of Urology} \bibinfo{volume}{189},
  \bibinfo{pages}{86--92}.
\bibitem[{Sung et~al.(2021)Sung, Ferlay, Siegel, Laversanne, Soerjomataram,
  Jemal and Bray}]{globalstat}
\bibinfo{author}{Sung, H.}, \bibinfo{author}{Ferlay, J.},
  \bibinfo{author}{Siegel, R.L.}, \bibinfo{author}{Laversanne, M.},
  \bibinfo{author}{Soerjomataram, I.}, \bibinfo{author}{Jemal, A.},
  \bibinfo{author}{Bray, F.}, \bibinfo{year}{2021}.
\newblock \bibinfo{title}{Global cancer statistics 2020: Globocan estimates of
  incidence and mortality worldwide for 36 cancers in 185 countries}.
\newblock \bibinfo{journal}{CA: A Cancer Journal for Clinicians}
  \bibinfo{volume}{71}, \bibinfo{pages}{209--249}.
\bibitem[{Tian et~al.(2020)Tian, Krishnan and Isola}]{tian2019crd}
\bibinfo{author}{Tian, Y.}, \bibinfo{author}{Krishnan, D.},
  \bibinfo{author}{Isola, P.}, \bibinfo{year}{2020}.
\newblock \bibinfo{title}{Contrastive representation distillation}, in:
  \bibinfo{booktitle}{International Conference on Learning Representations}.
\bibitem[{Tătaru et~al.(2021)Tătaru, Vartolomei, Rassweiler, Virgil,
  Lucarelli, Porpiglia, Amparore, Manfredi, Carrieri, Falagario, Terracciano,
  de~Cobelli, Busetto, Giudice and Ferro}]{diagnostics11020354}
\bibinfo{author}{Tătaru, O.S.}, \bibinfo{author}{Vartolomei, M.D.},
  \bibinfo{author}{Rassweiler, J.J.}, \bibinfo{author}{Virgil, O.},
  \bibinfo{author}{Lucarelli, G.}, \bibinfo{author}{Porpiglia, F.},
  \bibinfo{author}{Amparore, D.}, \bibinfo{author}{Manfredi, M.},
  \bibinfo{author}{Carrieri, G.}, \bibinfo{author}{Falagario, U.},
  \bibinfo{author}{Terracciano, D.}, \bibinfo{author}{de~Cobelli, O.},
  \bibinfo{author}{Busetto, G.M.}, \bibinfo{author}{Giudice, F.D.},
  \bibinfo{author}{Ferro, M.}, \bibinfo{year}{2021}.
\newblock \bibinfo{title}{Artificial intelligence and machine learning in
  prostate cancer patient management—current trends and future perspectives}.
\newblock \bibinfo{journal}{Diagnostics} \bibinfo{volume}{11}.
\bibitem[{Vesal et~al.(2021)Vesal, Gu, Maier and Ravikumar}]{vesal2020}
\bibinfo{author}{Vesal, S.}, \bibinfo{author}{Gu, M.}, \bibinfo{author}{Maier,
  A.}, \bibinfo{author}{Ravikumar, N.}, \bibinfo{year}{2021}.
\newblock \bibinfo{title}{Spatio-temporal multi-task learning for cardiac mri
  left ventricle quantification}.
\newblock \bibinfo{journal}{IEEE Journal of Biomedical and Health Informatics}
  \bibinfo{volume}{25}, \bibinfo{pages}{2698--2709}.
\bibitem[{Wang and Yoon(2022)}]{Lin2022}
\bibinfo{author}{Wang, L.}, \bibinfo{author}{Yoon, K.J.}, \bibinfo{year}{2022}.
\newblock \bibinfo{title}{Knowledge distillation and student-teacher learning
  for visual intelligence: A review and new outlooks}.
\newblock \bibinfo{journal}{IEEE Transactions on Pattern Analysis and Machine
  Intelligence} \bibinfo{volume}{44}, \bibinfo{pages}{3048--3068}.
\newblock \DOIprefix\doi{10.1109/TPAMI.2021.3055564}.
\bibitem[{Wang et~al.(2019)Wang, Dou, Hu, Zhu, Yang, Xu, Qin, Heng, Wang and
  Ni}]{wang2019}
\bibinfo{author}{Wang, Y.}, \bibinfo{author}{Dou, H.}, \bibinfo{author}{Hu,
  X.}, \bibinfo{author}{Zhu, L.}, \bibinfo{author}{Yang, X.},
  \bibinfo{author}{Xu, M.}, \bibinfo{author}{Qin, J.}, \bibinfo{author}{Heng,
  P.A.}, \bibinfo{author}{Wang, T.}, \bibinfo{author}{Ni, D.},
  \bibinfo{year}{2019}.
\newblock \bibinfo{title}{Deep attentive features for prostate segmentation in
  3d transrectal ultrasound}.
\newblock \bibinfo{journal}{IEEE Transactions on Medical Imaging}
  \bibinfo{volume}{38}, \bibinfo{pages}{2768--2778}.
\bibitem[{Wang and Ji(2018)}]{wang2018smoothed}
\bibinfo{author}{Wang, Z.}, \bibinfo{author}{Ji, S.}, \bibinfo{year}{2018}.
\newblock \bibinfo{title}{Smoothed dilated convolutions for improved dense
  prediction}, in: \bibinfo{booktitle}{Proceedings of the 24th ACM SIGKDD
  International Conference on Knowledge Discovery \& Data Mining},
  \bibinfo{organization}{ACM}. pp. \bibinfo{pages}{2486--2495}.
\bibitem[{Weiss et~al.(2016)Weiss, Khoshgoftaar and Wang}]{Weiss2016}
\bibinfo{author}{Weiss, K.}, \bibinfo{author}{Khoshgoftaar, T.M.},
  \bibinfo{author}{Wang, D.}, \bibinfo{year}{2016}.
\newblock \bibinfo{title}{A survey of transfer learning}.
\newblock \bibinfo{journal}{Journal of Big Data} \bibinfo{volume}{3},
  \bibinfo{pages}{9}.
\bibitem[{Williams et~al.(2022)Williams, Ahdoot, Daneshvar, Hague, Wilbur,
  Gomella, Shih, Khondakar, Yerram, Mehralivand, Gurram, Siddiqui, Pinsky,
  Parnes, Merino, Wood, Turkbey and Pinto}]{Williams2022}
\bibinfo{author}{Williams, C.}, \bibinfo{author}{Ahdoot, M.},
  \bibinfo{author}{Daneshvar, M.A.}, \bibinfo{author}{Hague, C.},
  \bibinfo{author}{Wilbur, A.R.}, \bibinfo{author}{Gomella, P.T.},
  \bibinfo{author}{Shih, J.}, \bibinfo{author}{Khondakar, N.},
  \bibinfo{author}{Yerram, N.}, \bibinfo{author}{Mehralivand, S.},
  \bibinfo{author}{Gurram, S.}, \bibinfo{author}{Siddiqui, M.},
  \bibinfo{author}{Pinsky, P.}, \bibinfo{author}{Parnes, H.},
  \bibinfo{author}{Merino, M.}, \bibinfo{author}{Wood, B.},
  \bibinfo{author}{Turkbey, B.}, \bibinfo{author}{Pinto, P.A.},
  \bibinfo{year}{2022}.
\newblock \bibinfo{title}{Why does magnetic resonance imaging-targeted biopsy
  miss clinically significant cancer?}
\newblock \bibinfo{journal}{Journal of Urology} \bibinfo{volume}{207},
  \bibinfo{pages}{95--107}.
\bibitem[{Xu et~al.(2021)Xu, Sanford, Turkbey, Xu, Wood and Yan}]{Xu2021}
\bibinfo{author}{Xu, X.}, \bibinfo{author}{Sanford, T.},
  \bibinfo{author}{Turkbey, B.}, \bibinfo{author}{Xu, S.},
  \bibinfo{author}{Wood, B.J.}, \bibinfo{author}{Yan, P.},
  \bibinfo{year}{2021}.
\newblock \bibinfo{title}{Shadow-consistent semi-supervised learning for
  prostate ultrasound segmentation}.
\newblock \bibinfo{journal}{IEEE Transactions on Medical Imaging} ,
  \bibinfo{pages}{1--1}.
\bibitem[{Yang et~al.(2016)Yang, Rossi, Jani, Mao, Curran and Liu}]{Yang2016}
\bibinfo{author}{Yang, X.}, \bibinfo{author}{Rossi, P.J.},
  \bibinfo{author}{Jani, A.B.}, \bibinfo{author}{Mao, H.},
  \bibinfo{author}{Curran, W.J.}, \bibinfo{author}{Liu, T.},
  \bibinfo{year}{2016}.
\newblock \bibinfo{title}{{3D transrectal ultrasound (TRUS) prostate
  segmentation based on optimal feature learning framework}}, in:
  \bibinfo{booktitle}{Medical Imaging 2016: Image Processing},
  \bibinfo{organization}{International Society for Optics and Photonics}.
  \bibinfo{publisher}{SPIE}. pp. \bibinfo{pages}{654 -- 660}.
\bibitem[{Yang et~al.(2017)Yang, Yu, Wu, Wang, Ni, Qin and Heng}]{yang2017}
\bibinfo{author}{Yang, X.}, \bibinfo{author}{Yu, L.}, \bibinfo{author}{Wu, L.},
  \bibinfo{author}{Wang, Y.}, \bibinfo{author}{Ni, D.}, \bibinfo{author}{Qin,
  J.}, \bibinfo{author}{Heng, P.A.}, \bibinfo{year}{2017}.
\newblock \bibinfo{title}{Fine-grained recurrent neural networks for automatic
  prostate segmentation in ultrasound images}, in:
  \bibinfo{booktitle}{Proceedings of the Thirty-First AAAI Conference on
  Artificial Intelligence}, \bibinfo{publisher}{AAAI Press}. p.
  \bibinfo{pages}{1633–1639}.
\bibitem[{Yun et~al.(2020)Yun, Park, Lee and Shin}]{Yun_2020_CVPR}
\bibinfo{author}{Yun, S.}, \bibinfo{author}{Park, J.}, \bibinfo{author}{Lee,
  K.}, \bibinfo{author}{Shin, J.}, \bibinfo{year}{2020}.
\newblock \bibinfo{title}{Regularizing class-wise predictions via
  self-knowledge distillation}, in: \bibinfo{booktitle}{The IEEE/CVF Conference
  on Computer Vision and Pattern Recognition (CVPR)}.
\bibitem[{Zeng et~al.(2018)Zeng, Samei, Karimi, Kesch, Mahdavi, Abolmaesumi and
  Salcudean}]{Zeng2018}
\bibinfo{author}{Zeng, Q.}, \bibinfo{author}{Samei, G.},
  \bibinfo{author}{Karimi, D.}, \bibinfo{author}{Kesch, C.},
  \bibinfo{author}{Mahdavi, S.S.}, \bibinfo{author}{Abolmaesumi, P.},
  \bibinfo{author}{Salcudean, S.E.}, \bibinfo{year}{2018}.
\newblock \bibinfo{title}{Prostate segmentation in transrectal ultrasound using
  magnetic resonance imaging priors}.
\newblock \bibinfo{journal}{International Journal of Computer Assisted
  Radiology and Surgery} \bibinfo{volume}{13}, \bibinfo{pages}{749--757}.
\bibitem[{Zhan and Shen(2006)}]{Zhan2006}
\bibinfo{author}{Zhan, Y.}, \bibinfo{author}{Shen, D.}, \bibinfo{year}{2006}.
\newblock \bibinfo{title}{Deformable segmentation of 3-d ultrasound prostate
  images using statistical texture matching method}.
\newblock \bibinfo{journal}{IEEE Transactions on Medical Imaging}
  \bibinfo{volume}{25}, \bibinfo{pages}{256--272}.
\bibitem[{Zhou et~al.(2020)Zhou, Yang, Hospedales and Xiang}]{zhou2020}
\bibinfo{author}{Zhou, K.}, \bibinfo{author}{Yang, Y.},
  \bibinfo{author}{Hospedales, T.}, \bibinfo{author}{Xiang, T.},
  \bibinfo{year}{2020}.
\newblock \bibinfo{title}{Learning to generate novel domains for domain
  generalization}, in: \bibinfo{editor}{Vedaldi, A.}, \bibinfo{editor}{Bischof,
  H.}, \bibinfo{editor}{Brox, T.}, \bibinfo{editor}{Frahm, J.M.} (Eds.),
  \bibinfo{booktitle}{Computer Vision -- ECCV 2020}, pp.
  \bibinfo{pages}{561--578}.
\bibitem[{Zhou et~al.(2018)Zhou, Rahman~Siddiquee, Tajbakhsh and
  Liang}]{nestedUnet2018}
\bibinfo{author}{Zhou, Z.}, \bibinfo{author}{Rahman~Siddiquee, M.M.},
  \bibinfo{author}{Tajbakhsh, N.}, \bibinfo{author}{Liang, J.},
  \bibinfo{year}{2018}.
\newblock \bibinfo{title}{Unet++: A nested u-net architecture for medical image
  segmentation}, in: \bibinfo{booktitle}{Deep Learning in Medical Image
  Analysis and Multimodal Learning for Clinical Decision Support}, pp.
  \bibinfo{pages}{3--11}.

\end{thebibliography}

\section*{Supplementary Material}
\beginsupplement

\subsection*{Quantitative Evaluation}
\begin{figure}[htbp]
    \centering
    \includegraphics[width=0.45\textwidth]{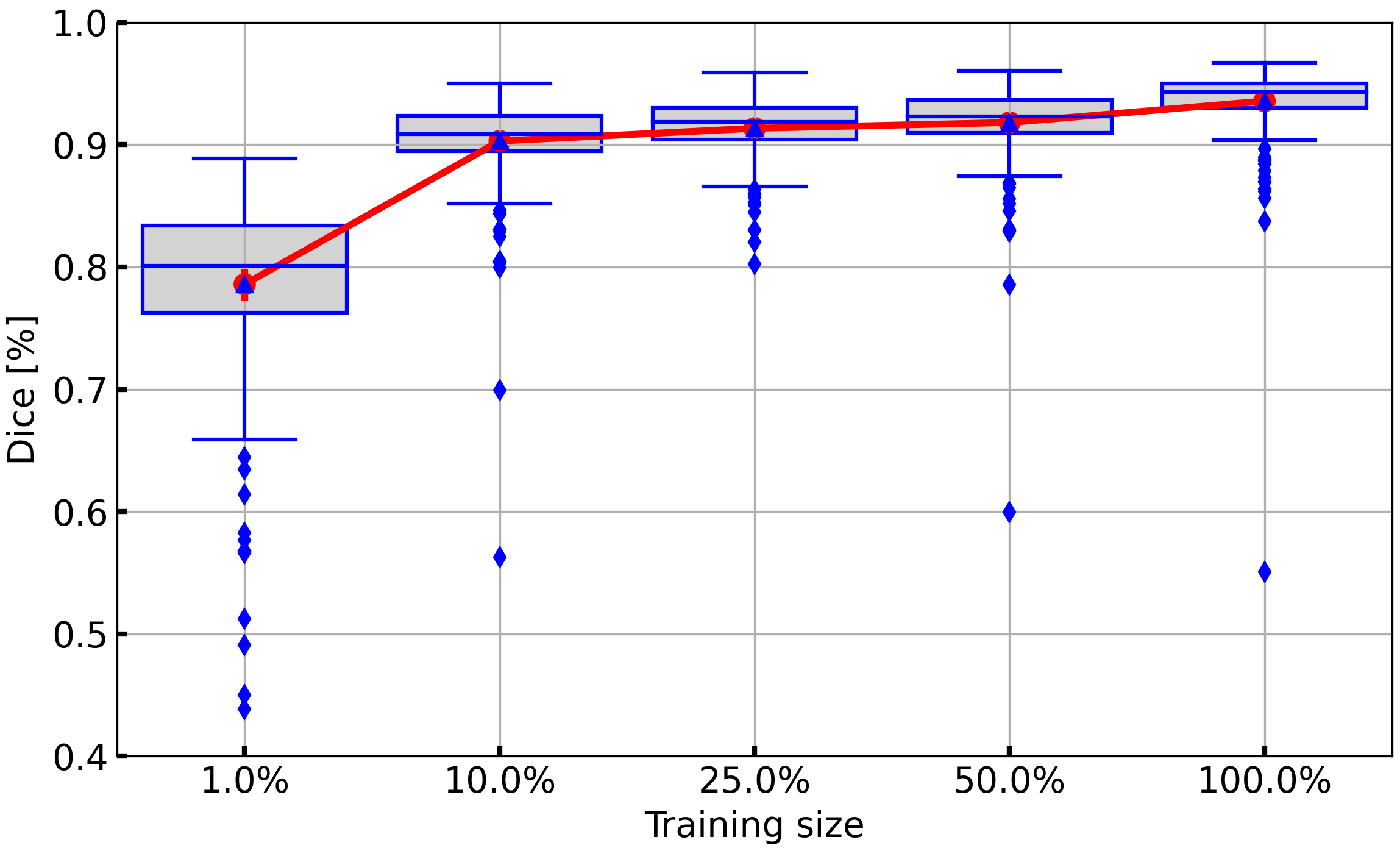}
    \caption{Boxplot of Dice scores achieved by CoordDR-UNet with different training sizes on cohort C1-test.}
    \label{fig:trainingsize}
\end{figure}

To evaluate the impact of training data size with respect to segmentation accuracy, we trained our 2.5D CoordDR-UNet model with different percentages of training data. The model achieved a Dice score of 0.78 on 220 TRUS images from cohort C1-test using only one percent of the training data (Fig. \ref{fig:trainingsize}). The Dice score increased to 0.90 with only ten percent of training data (80 TRUS images) and 0.94 with a hundred percent of the data. Noise in the annotations by the urologist in the training data prevented further model convergence. Table \ref{tab:S1} shows the quantitative results for different models after finetuning using the CoordDR-UNet + KDL model. Overall, the model $\mathbf{M}_{3}$ achieved the best average Dice score across all three datasets.   

\begin{table}[htbp]
    \centering
    \caption{Quantitative comparison results $(mean (\pm std))$ for different models. M1: model trained on cohort C1, M2: model finetuned on cohort C2, and M3: model finetuned on cohort C2 and then on cohort C3. Bolded values show the best results column-wise.}
    \label{tab:S1}
    \resizebox{0.48\textwidth}{!}{%
    \begin{tabular}{l|c|c|c|c}
    \hline
    
    \textbf{Models} & \begin{tabular}[c]{@{}l@{}} \textbf{Cohort C1} \\ \textbf{(In-house)}\end{tabular} & \begin{tabular}[c]{@{}l@{}} \textbf{Cohort C2} \\  \textbf{(Public)}\end{tabular} & \begin{tabular}[c]{@{}l@{}} \textbf{Cohort C3} \\ \textbf{(External)}\end{tabular} & \textbf{Average} \\ \hline
    $\mathbf{M}_{1}$ & \textbf{0.94 ($\pm$0.03)} & 0.89 ($\pm$0.03) & 0.43 ($\pm$0.31) & 0.75 ($\pm$0.13)\\ \hline
    $\mathbf{M}_{2}$ & 0.93 ($\pm$0.03) & \textbf{0.91 ($\pm$0.03)} & 0.24 ($\pm$0.29) & 0.69 ($\pm$0.12)\\ \hline
    $\mathbf{M}_{3}$ & 0.91 ($\pm$0.04) & 0.88 ($\pm$0.03) & \textbf{0.82 ($\pm$0.16)} & \textbf{0.87 ($\pm$0.08)}\\ \hline
    \end{tabular}%
    }
\end{table}

\subsection*{Model hyperparemters}
Table \ref{tab:S2} shows the training hyperparameters used for all the 2D, 2.5D, and 3D models in this study including UNet, Attention-UNet, Nested-UNet, Dilated-residual UNet, DAFNet, and CoordDR-UNet + LKD.
\begin{table}[!htbp]
    \centering
    \caption{Hyperparameters used for all models trained for prostate gland segmentation experiments.}
    \label{tab:S2}
    \resizebox{0.48\textwidth}{!}{%
    \begin{tabular}{l|ccc}
    \hline
    \textbf{Hyperparameter} & \multicolumn{1}{c}{\textbf{2D Models}} & \multicolumn{1}{c}{\textbf{2.5D Models}} & \multicolumn{1}{c}{\textbf{3D Models}} \\ \hline
    Input size        & $128 \times 160$ & $128 \times 160 \times 3$ & $128 \times 160 \times 80$ \\\hline
    Optimizer         & \multicolumn{3}{c}{Adam ($\beta_{1} = 0.9$ and $\beta_{2}=0.999$)}       \\ \hline
    Weight decay ($\alpha$)      & $0.01 $ & $0.01$ & $0.01 $                                       \\
    Loss function     & Soft-Dice loss  & Soft-Dice loss& Soft-Dice loss                                      \\
    Learning rate ($\eta$)     & $10^{-3}$  & $10^{-3}$ & $10^{-3}$                                    \\
    LR decay schedule & StepLR   & StepLR & StepLR                                             \\
    Train epochs      & 500        & 500  & 200                                           \\
    Batch size        & 64               & 64                        & 1                         \\
    Data augmentation & Yes              & Yes                       & Yes                       \\ \hline
    \end{tabular}%
    }
\end{table}
\end{document}